\documentclass[prd,superscriptaddress,showpacs,onecolumn]{revtex4-2}
\usepackage{bm}
\usepackage{amsfonts}
\usepackage{latexsym}
\usepackage[latin1]{inputenc}
\usepackage{graphicx}
\usepackage{amsmath}
\usepackage{palatino}
\usepackage{mathpazo}
\usepackage{textcomp}
\usepackage{float}
\usepackage{booktabs}
\usepackage{dcolumn}
\usepackage{hyperref}
\usepackage{amsmath}
\usepackage{xcolor}
\usepackage{orcidlink}
\usepackage[caption=false]{subfig}
\usepackage{commath}

\hypersetup{colorlinks,citecolor=blue}
\def\jnl@style{\it}
\def\aaref@jnl#1{{\jnl@style#1}}
\def\aaref@jnl#1{{\jnl@style#1}}
\def\aj{\aaref@jnl{AJ}}
\def\apj{\aaref@jnl{ApJ}}
\def\apjl{\aaref@jnl{ApJ}}
\def\apjs{\aaref@jnl{ApJS}}
\def\apss{\aaref@jnl{Ap\&SS}}
\def\aap{\aaref@jnl{A\&A}}
\def\aapr{\aaref@jnl{A\&A~Rev.}}
\def\aaps{\aaref@jnl{A\&AS}}
\def\mnras{\aaref@jnl{Mon.~Not.~Roy.~Astron.~Soc.}}
\def\prd{\aaref@jnl{Phys.~Rev.~D}}
\def\prc{\aaref@jnl{Phys.~Rev.~C}}
\def\prl{\aaref@jnl{Phys.~Rev.~Lett.}}
\def\qjras{\aaref@jnl{QJRAS}}
\def\skytel{\aaref@jnl{S\&T}}
\def\ssr{\aaref@jnl{Space~Sci.~Rev.}}
\def\zap{\aaref@jnl{ZAp}}
\def\nat{\aaref@jnl{Nature}}
\def\aplett{\aaref@jnl{Astrophys.~Lett.}}
\def\apspr{\aaref@jnl{Astrophys.~Space~Phys.~Res.}}
\def\physrep{\aaref@jnl{Phys.~Rep.}}
\def\physscr{\aaref@jnl{Phys.~Scr}}
\def\commat{\aaref@jnl{Comm.~Math.~Phys.}}
\def\science{\aaref@jnl{Science}}
\def\cqg{\aaref@jnl{Classical Quant.~Grav.}}
\def\jpcs{\aaref@jnl{JPCS}}
\def\ijmpd{\aaref@jnl{Int.~J.~Mod.~Phys.~D}}
\def\grg{\aaref@jnl{Gen.~Relat.~Gravit.}}
\def\rpp{\aaref@jnl{Rep.~Prog.~Phys.}}
\def\npa{\aaref@jnl{Nucl.~Phys.~A}}
\def\lrr{\aaref@jnl{Living Rev.~Rel.}}
\def\jcap{\aaref@jnl{J.~Cosmology Astropart.~Phys.}}
\def\rmp{\aaref@jnl{Rev.~Mod.~Phys.}}
\def\epjc{\aaref@jnl{Eur.~Phys.~J.~C}}

\allowdisplaybreaks[1]
\renewcommand{\arraystretch}{1.1}
\begin{document}

\title{Reconstructing cosmic evolution with a density parametrization}
\author{Ritika Nagpal\orcidlink{0000-0002-4669-6395}}
\email{ritikanagpal.math@gmail.com}
\author{Shibesh Kumar Jas Pacif\orcidlink{0000-0003-0951-414X}}
\email{shibesh.math@gmail.com}
\author{Abhishek Parida\orcidlink{0000-0000-0000-0000}}
\email{abhishekparida22@gmail.com}
\date{\today }

\begin{abstract}
The current paper provides a comprehensive examination of a dark energy
cosmological model in the classical regime, in which a generic scalar field
is regarded as a dark energy source. Einstein's field equations are solved
in model independent way i.e. using a scheme of cosmological
parametrization. A parametrization of the density parameter as a function of
the cosmic scale factor has been investigated in this line. The result is
noteworthy because it shows a smooth transition from a decelerating to an
accelerating phase in the recent past. The model parameters involved in the
functional form of the parametrization approach utilized here were
constrained using certain external datasets. The updated Hubble datasets
containing 57 datapoints, 1048 points of recently compiled Pantheon
datasets, and also the Baryon Acoustic Oscillation (BAO) datasets are used
here to determine the best-fitting model parameter values. The expressions
of several significant cosmological parameters are represented as a function
of redshift `$z$' and illustrated visually for the best fit values of the
model parameters to better comprehend cosmic evolution. The obtained model
is also compared with the $\Lambda CDM$ model. Our model has a distinct
behavior in future and shown a big crunch type collapse. The best fit values
of the model parameters are also used to compute the current values of
several physical and geometrical parameters, as well as phase transition
redshift. To examine the nature of dark energy, certain cosmological tests
and diagnostic analyses are done on the derived model.
\end{abstract}

\maketitle

\color{black} 

\affiliation{Department of Mathematics, Vivekananda College, University of
Delhi, New Delhi, Delhi 110032, India.}

\affiliation{Centre for Cosmology and Science Popularization (CCSP), SGT
University,	Gurugram, Delhi-NCR, Gurugram 122505, Haryana, India.}

\affiliation{International College of Liberal Arts, Yamanashi Gakuin
University,	Yamanashi 400-0805, Japan.}


\section{Introduction}

\label{sec1}

Before $1916$, gravity was assumed to be an intrinsic property of objects 
\textit{- }a constant instantaneous force that could act over long
distances. But, the intriguing discovery of Einstein's theory of general
relativity (GR) has changed the course of scientific history. GR solved the
riddle of mercury's precision and explained that gravity was not a
mysterious force acting at a distance in the background of space and time
but resulted from bending the background itself. The key insight of the GR
is that the shortest path between any two distant objects in the space is
always curved. This curved geometry is the basis of GR. Scientist's paradigm
about the cosmos has wholly changed after Einstein's theory of gravity over
the past $100$ years. Einstein's field equations (EFEs) have expressed many
things analytically, and somehow this theory matched the observations, which
had been a mystery for decades. Many skepticisms still present after $100$
years of Einstein's GR, such as its inability to explain the singularities
inside the black holes, big bang, age of the Universe and few more \cite%
{GRProblems2016}. One of GR's biggest challenges is, understanding the
curvature singularities, geodesic incompleteness, and b-incompleteness. Many
inherent features of the Universe have led to plenty of consequences and
conjecture in the field of GR. Therefore, one of the major pursuits in
physics is finding a better theory. The Einstein's theory axed with
theoretical as well as observational issues several times in the past
century. However, new gravitational wave findings and a black hole image
enhance GR's foundation. So, we are interested here to explore the late
evolution of the Universe in the background of GR.

The present scenario with the expanding Universe is that it has gone \
through a series of evolution processes after the big bang where the two
phases of accelerating expansion of the Universe play some major role. The
first phase of cosmic acceleration is known as \textit{inflation} (that
occurred just after the big bang) and the second phase is the recently
discovered \textit{late-time acceleration} \cite{Reiss1998}, \cite%
{Perlmutter1999}, \cite{Tegamark2004}, \cite{Hinshaw2013}. The detection of
Type Ia Supernovae (SNeIa) \cite{Reiss1998} provided the first evidence for
cosmic acceleration in 1998. Afterward, many other observations have
revealed strong support for the cosmic acceleration at late-times \cite%
{Rose2020} and many others support this fact indirectly \cite{Jaffe2000}, 
\cite{CMB2007}\thinspace\ \cite{BAO2006}, \cite{SELJAK2004}, \cite{SDSS2006}%
. This cosmic speed-up bring about either the finding for a mysterious
energy content of the space-time or for modifications in GR that could
confront with such a concealed feature of the Universe. A negative pressure
fluid must be introduced into the Universe's energy content to retain GR
with acceleration in the classical regime. The term cosmological constant $%
\Lambda $ (a tiny number $\approx 1.3\ast 10^{-52}m^{-2}$) into the Einstein
field equations describes the intrinsic energy density of the vacuum, which
plays the role of the most intriguing candidate of dark energy (DE).
However, the mathematical expression $\Lambda $ edges to a massive variation
between the theoretical and observation predictions \cite{Weinberg1989}.
Many DE models \cite{Steinhardt1999} have been created as a result of the
variety caused by the issue of fine-tuning and the cosmic coincidence
problem related with CDM. A phenomenological solution of these issues with
the standard model can be figured out by considering a cosmological term
varying with time \cite{Wetterich1995}, \cite{Peebles1988}. Thus, the scalar
field models have gained high popularity in recent decades as they play a
crucial role in describing both early and late time cosmic acceleration by
acting as a DE candidate \cite{Copeland2006}. The motivation of interest in
this paper is to develop a cosmological model in terms of a scalar field
that unifies the characteristic of scalar field cosmologies. A dynamically
evolving scalar field can be utilized as a candidate of dark energy. A
cosmological constant can be a slow roll scalar field. Other dark energy
models with a general scalar field include quintessence \cite{quint1}, \cite%
{quint2}, \cite{quint3}, phantom \cite{phant1}, \cite{phant2}, \cite{phant3}%
, \cite{phant4}, K-essence \cite{kessen1}, \cite{kessen2}, \cite{kessen3}, 
\cite{kessen4}, quintom \cite{quintom1}, \cite{quintom2}, \cite{quintom3}
etc., which can explain the Universe's evolution nicely. Few other
candidates are tachyon scalar field \cite{tach1}, \cite{tach2}, \cite{tach3}
chameleon \cite{chameleon1}, \cite{chameleon2}, \cite{chameleon3} scalar
field, \cite{galileon1}, \cite{galileon2}, \cite{galileon3} etc. The theory
is that the field must exert a lot of negative pressure in order to speed up
the expansion of the Universe. The exotic fluid such as Chapligyn gas \cite%
{gorini2003}, Polytropic gas \cite{chavanis2012} can act as the candidates
of dark energy. However, we are discussing a model in a model-independent
way and are using a generic scalar field.

In the literature, the model-independent technique (or cosmological
parametrization) of reconstructing a cosmological model with or without dark
energy has been employed to fit data to the cosmic evolution of the
Universe. The model independent approach in the framework of some DE
candidates is of great interest nowadays and was first discussed by
Starobinsky \cite{Staro1998}. A broad variety of parametrization schemes 
\cite{SKJPACIF} have been proposed in the literature to explain the
development of the Universe, including the shift from early deceleration to
late acceleration. Moreover, there are other parametrization schemes such as
density parametrization, pressure parametrization together with the
parametrization of deceleration parameter, Hubble parameter, Scale factor
and more (For a detailed list of parametrization schemes see \cite{SKJPACIF}%
). Therefore, the purpose of this paper is to advocate for a particular
parametrization of the scalar field's energy density, which better explains
cosmic dynamics and provides more compact limitations than any other
cosmological parameter.

This paper's work has been arranged as follows: Section I gives a quick
overview of general relativity and the present state of some of the most
pressing cosmological issues. The basic set up of scalar field as a
candidate for dark energy in GR is explored in Sect. II. In Sect. III, we
consider the space-time metric and formulated the Einstein Field equations
in GR. We got the solutions to field equations using a basic parametrization
approach in Sect. IV. In Sect. V, we have constrained the model parameters
involved in our model using some external datasets and found the best fit
values of some cosmological parameters. In Sect. VI, some detailed analysis
of the model through the behavior of physical parameters through graphical
representation are discussed. We examined the feasibility of our model via
energy conditions in Sect. VII, and we wrap up our findings in Sect. VIII.

\section{Basic Formalism with a general Scalar field}

\label{sec2}

Since, dark energy has been a mystery until now and because of a lack of
understanding of the matter sector, selecting a suitable candidate for dark
energy is a difficult undertaking. However, it is generally represented by a
x-matter or a large-scale scalar field $\phi $. So, in this paper, we will
examine an ordinary scalar field with the Lagrangian density $L=-\frac{1}{2}%
g^{\mu \nu }\partial _{\mu }\phi \partial _{\nu }\phi -V(\phi )$ that is
minimally related to gravity (technically known as quintessence field), then
the gravitational action for the scalar field will be given by,

\begin{equation}
S=\int d^{4}x\sqrt{-g}\left[ -\frac{1}{2}g^{\mu \nu }\partial _{\mu }\phi
\partial _{\nu }\phi -V\left( \phi \right) \right] ,  \label{ACTION}
\end{equation}%
where $V\left( \phi \right) $ denotes the scalar field's potential. Because
dark energy is supposed to be homogenous, we may consider the scalar field
to be spatially homogeneous, in which case the scalar field's stress-energy
tensor also assumes the form of a perfect fluid and can be expressed as,

\begin{equation}
T_{\mu \nu }^{\phi }=T_{\mu \nu }^{Dark~Energy}=\left( \rho
_{\phi}+p_{\phi}\right) U_{\mu }U_{\nu }+p_{\phi }g_{\mu \nu }.  \label{EMDE}
\end{equation}

Where $\rho _{\phi }$ and $p_{\phi }$ respectively represent the scalar
field's energy density and pressure. The wave equation governs the temporal
development of the scalar field and is given by,%
\begin{equation}
\ddot{\phi}+3H\dot{\phi}+\frac{dV(\phi )}{d\phi }=0.  \label{KGEQ}
\end{equation}

A derivative with regard to time $t$ is represented by an overhead dot from
now on.

Einstein's gravitational field equations are as follows:

\begin{equation}
R_{\mu \nu }-\frac{1}{2}Rg_{\mu \nu }=-8\pi GT_{\mu \nu },  \label{FE}
\end{equation}%
with $c=1$, where $T_{\mu \nu }=(\rho +p)U_{\mu }U_{\nu }+p~g_{\mu \nu }$
represents the Universe's ideal source of fluid matter (perfect fluid). With
the introduction of dark energy into the Einstein field equations, the
energy momentum tensor $T_{\mu \nu }$ will be modified to $T_{\mu \nu
}^{Total}$, where

\begin{equation}
T_{\mu \nu }^{Total}=T_{\mu \nu }^{Matter}+T_{\mu \nu }^{Dark~Energy}=\left(
\rho _{Total}+p_{Total}\right) U_{\mu }U_{\nu }+p_{Total}g_{\mu \nu }
\label{TOEM}
\end{equation}

By considering the low contribution of baryons and radiation, we write $\rho
_{Total}=\rho _{m}+\rho _{\phi }$ and $p_{Total}=p_{m}+p_{\phi }$. Here, $%
\rho _{m}$ and $p_{m}$ are respectively, the energy density and pressure of
the dark matter in the Universe. Since the dark matter pressure is
negligible, we take $p_{m}\approx 0$. In the next section, we shall apply
this set up to a specific metric space and find the solution of EFEs.

\section{Einstein Field Equations}

\label{sec3}

To begin, let us suppose that the Universe is homogenous and isotropic. So,
we will use the Friedmann-Lemaitre-Robertson-Walker space-time with flat
geometry as the backdrop metric in the form,

\begin{equation}
ds^{2}=dt^{2}-a^{2}(t)(dx^{2}+dy^{2}+dz^{2}),  \label{1rs}
\end{equation}%
$a(t)$ denotes the Universe's scale factor. With the above matter source
description in the Universe, the Friedmann equations can be represented in
the form,

\begin{equation}
\rho _{Total}=3M_{pl}^{2}H^{2},  \label{2rs}
\end{equation}%
\begin{equation}
\rho _{Total}+3p_{Total}=6M_{pl}^{2}H^{2}q,  \label{3rs}
\end{equation}%
where, $H\left( =\frac{\dot{a}}{a}\right) $ and $q\left( =-\frac{a\ddot{a}}{ 
\dot{a}^{2}}\right) $ are Hubble parameter and deceleration parameter
respectively and $M_{pl}=\left( 8\pi G\right) ^{-1/2}$ is the Planck mass.
If we consider the effective equation of state as $p_{Total}=\omega
_{Total}\rho _{Total}$ then it is easy to write,

\begin{equation}
q=\frac{1}{2}\left( 1+3\omega _{Total}\right)  \label{4rs}
\end{equation}

From equations Eq. (\ref{2rs}) and Eq. (\ref{3rs}), the total conservation
equation can be derived as,

\begin{equation}
\dot{\rho}_{Total}+3H\left( \rho _{Total}+p_{Total}\right) =0  \label{5rs}
\end{equation}%
\qquad

Two scenario comes into picture now. Either matter field and scalar field
interact gravitationally or have minimal interaction. Let us consider the
interaction between the two fields is minimal which leads to

\begin{eqnarray}
\dot{\rho}_{m}+3H\left( \rho _{m}+p_{m}\right) &=&0\,  \label{6rs} \\
\,\dot{\rho}_{\phi }+3H\left( \rho _{\phi }+p_{\phi }\right) &=&0
\label{6ars}
\end{eqnarray}

Since, the dark matter pressure is negligible, we have $p_{m}=0$, the above
equation Eq. (\ref{6rs}) yield $\rho _{m}=c_{1}a^{-3}$, where $c_{1}$ is a
constant of integration.

The energy density and pressure for the scalar field under consideration in
the FLRW backdrop may be expressed as,

\begin{equation}
\rho _{\phi }=\frac{\dot{\phi}^{2}}{2}+V\left( \phi \right) \text{, }p_{\phi
}=\frac{\dot{\phi}^{2}}{2}-V\left( \phi \right) .  \label{6brs}
\end{equation}

Here, $\frac{\dot{\phi}^{2}}{2}$ and $V\left( \phi \right) $ are
respectively the kinetic term and potential term for the field respectively.
If the potential term dominates over the kinetic term i.e. $\dot{\phi}%
^{2}<<V(\phi )$ (slow roll scalar field), then the equation of state
parameter $w_{\phi }$,

\begin{equation}
w_{\phi }=\frac{\frac{\dot{\phi}^{2}}{2}-V\left( \phi \right) }{\frac{\dot{%
\phi}^{2}}{2}+V\left( \phi \right) }=\frac{-1+\frac{\dot{\phi}^{2}}{2V}}{1+%
\frac{\dot{\phi}^{2}}{2V}}  \label{6crs}
\end{equation}%
resumes the value to $w_{\phi }=-1$ and then the potential term would behave
like a cosmological constant $\Lambda $. we must mention here that, most
favoured candidate of dark energy is the Einstein's cosmological constant $%
\Lambda $ and $\Lambda $CDM model has the nice fits to most of the
observational datasets. However, for the dynamical case, the values of $%
w_{\phi }$ ranges between $-1$ and $+1$ and discuss some other candidates of
DE (see \cite{Copeland2006}). In the next section, we shall discuss the
solution in case of dynamical $w_{\phi }$.

\section{Solution of Field Equations}

\label{sec4}

From equation Eq. (\ref{6ars}), we can write

\begin{equation}
\rho _{\phi }=c_{2}\exp \left[ -3\int H\left( 1+\omega _{\phi }\right) dt%
\right] \text{,}  \label{7rs}
\end{equation}%
where $c_{2}$ is a constant of integration. For if $\omega _{\phi }=constant$%
, a particular case of slow roll scalar field (cosmological constant), Eq. (%
\ref{7rs}) yield $\rho _{\phi }=c_{2}a^{-3\left( 1+\omega _{\phi }\right)
}=constant$. But, we are considering here the dynamic equation of state for
the scalar field, i.e. $\omega _{\phi }\neq constant$ that serves as a
candidate of dynamical dark energy ($\phi $ varies with time $t$) and
describes the late time cosmic acceleration. In this scenario, the scalar
field $\phi $ is the only source of dark energy having potential $V(\phi )$
that interacts with itself. Eq. (\ref{7rs}) can also be represented as, 
\begin{equation}
\omega _{\phi }=-1-\frac{1}{3}a\frac{{\rho }_{\phi }^{^{\prime }}}{\rho
_{\phi }},  \label{8rs}
\end{equation}%
where the prime denotes the derivative \textit{w.r.t.} the scale factor $a$.
In order to solve above system of equations, which contains only three
independent equations with four unknowns $a$, $\rho _{m}$, $\rho _{\phi }$
and $p_{\phi }$, we need an additional constraint equation. As discussed in
the introduction about the model independent way or the cosmological
parametrization scheme, here we assume an appropriate parametrization of $%
\rho _{\phi }$ of the form,%
\begin{equation}
{\rho _{\phi }}=(a)^{\beta }\left[ cosh^{-1}\left( \frac{1}{\beta \ast a}%
\right) \right] ^{\alpha }  \label{9rs}
\end{equation}%
where $\alpha \in (0,1)$ and $\beta \in (0,1)$ are two model parameters
which can be constrained from some observational datasets.

The selection of $\rho _{\phi }$ is made in such a way that the term $%
cosh^{-1}\left( \frac{1+z}{\beta }\right) $ can generate a signature flip
behavior of equation of state parameter. Thus, the equation in (\ref{9rs})
indicates a concave downward movement (i.e. the gradient decreases at each
point over time $t$). There are various mathematical forms of these types of
parametrization in literature (see \cite{SKJPACIF}, where different
parametrization schemes used in the past few decades are summarized in
detail), which help to study different dark energy models, which are capable
of explaining some features of the observational cosmology. The most
interesting characteristic of this model independent way study is that, it
does not presuppose or influence the validity of gravitational theory nor
they violate the physical and geometric properties \cite{Weinberg1989}
rather it provides some satisfactory explanation to different scenarios
coming from observation and reconstruct the evolution of the Universe. The
important point in any parametrization scheme is that the new functional
form assumed contain some free parameters (model parameters), which should
be chosen properly and can be done by constraining the model with some
observational datasets. For this, it is better to write the expressions of
cosmological parameters in terms of redshift $z$. For a detailed study of
these kinds of model building/ reconstructions, one can refer to some recent
papers \cite{CP1}, \cite{CP2}, \cite{CP3}, \cite{CP4}, \cite{CP5}, \cite{CP6}%
, \cite{CP7}, \cite{CP8} in different scenarios.

Using the relation of redshift $z$ and scale factor $a$ given by $\frac{a}{%
a_{0}}=\frac{1}{(1+z)}$, where $a_{0}$ is the present ($t_{0}$ or $z=0$)
value of scale factor and normalized to $a_{0}=1$. We have then $\rho
_{m}=\rho _{m0}\left( 1+z\right) ^{3}$, (where $\rho _{m0}=\rho
_{m}(z=0)=\rho _{m}(t_{0})$ is the current value of the energy density of
dark matter). Now, the energy density of scalar field $\rho _{\phi }$ in
terms of redshift reads as, 
\begin{equation}
{\rho _{\phi }(z)}=(1+z)^{-\beta }\left\{ cosh^{-1}\left( \frac{1+z}{\beta }%
\right) \right\} ^{\alpha },  \label{10rs}
\end{equation}%
with $\rho _{\phi 0}=\rho _{\phi 0}(z=0)=[cosh^{-1}\left( \frac{1}{\beta }%
\right) ]^{\alpha }$ being the current value of the energy density of scalar
field and we can write $\rho _{\phi }(z)$ as, 
\begin{equation}
\rho _{\phi }(z)=\rho _{\phi 0}(1+z)^{-\beta }\frac{\left\{ cosh^{-1}\left( 
\frac{1+z}{\beta }\right) \right\} ^{\alpha }}{\left\{ cosh^{-1}\left( \frac{%
1}{\beta }\right) \right\} ^{\alpha }}  \label{12rs}
\end{equation}

Further, the Friedmann equation Eq. (\ref{2rs}) can be written as,

\begin{equation}
3M_{pl}^{2}H^{2}(z)=\rho _{m0}(1+z)^{3}+\rho _{\phi 0}\left\{
cosh^{-1}\left( \frac{1}{\beta }\right) \right\} ^{-\alpha }(1+z)^{-\beta
}\left\{ cosh^{-1}\left( \frac{1+z}{\beta }\right) \right\} ^{\alpha }
\label{13rs}
\end{equation}

Let us now introduce the density parameter $\Omega =\frac{\rho }{\rho _{c}}$%
, which describe the whole content of the Universe, where $\rho _{c}$ is the
critical density of the Universe and $\rho _{c}=\frac{3H^{2}}{8\pi G}%
=3M_{pl}^{2}H^{2}$. \newline

Equation Eq. (\ref{13rs}) in terms of density parameter of matter and scalar
field can be expressed as,

\begin{equation}
H(z)=H_{0}\left[ \Omega _{m0}(1+z)^{3}+\Omega _{\phi 0}\left\{
cosh^{-1}\left( \frac{1}{\beta }\right) \right\} ^{-\alpha }(1+z)^{-\beta
}\left\{ cosh^{-1}\left( \frac{1+z}{\beta }\right) \right\} ^{\alpha }\right]
^{\frac{1}{2}},  \label{14rs}
\end{equation}

where $\Omega _{m0}=\frac{\rho _{M0}}{3M_{pl}^{2}H_{0}^{2}}$ and $\Omega
_{\phi 0}=\frac{\rho _{\phi 0}}{3M_{pl}^{2}H_{0}^{2}}$ are the present
values of matter and scalar field density parameters respectively.\newline

The obtained model is described by the Eq. (\ref{14rs}) with the model
parameters $\alpha $ \& $\beta $ and the density parameters $\Omega _{m0}$
\& $\Omega _{\phi 0}$ which can be constrained using some observational
datasets. In the next section, we have constrained these model parameters
using some available external datasets such as Hubble datasets, Pantheon
datasets and Baryon Acoustic oscillation datasets and found the best fit
values of them for further analysis and discussed the behavior of other
physical and geometric parameters of the model.

\section{Statistical analysis of the model parameters}

\label{sec5}

The advancement in observational cosmology allow us to understand the
ancient and late cosmic evolution, the properties of dark components in the
Universe along with the structure formation. The past three decades of
cosmic studies (following Hubble's space telescope) in observational
cosmology have yielded an enormous amount of observational datasets, such as
SDSS, which produces a map of galaxy distribution and encodes current
variations in the Universe, CMBR, which verifies the big bang theory,
QUASARS, which clarifies the matter between observer and quasars, and BAO,
which estimates large scale structures in the Universe to better interpret
the DE. And also the SNeIa observations (known as standard candles) which
are the devices for computing the cosmic distances and many more. So, in
this part, we use error bar plots of Hubble datasets and $SNeIa$ Pantheon
datasets to compare our model to the $\Lambda CDM$ model, and we use
statistical analysis to restrict the values of model parameter \& included
in our model using $Hubble$, $Pantheon$, and $BAO$ datasets.

In this paper, we advocated for using Python's scipy optimization technique
to restrict the value of model parameters and anticipate the global minima
of Hubble function mentioned in Eq. (\ref{14rs}). The impressive
fluctuations within the inclining diagonal components of the covariance
matrix relating to the parameters are noticed. By employing the aforesaid
measurements as means and a Gaussian prior with a fixed value $\sigma =1.0$
as the dispersion, we used Python's emcee module for the mathematical
research and numerical analysis. Thus, we examined the parameter space
encompassing the local minima. The results are displayed as two-dimensional
contour plots with $1-\sigma $ and $2-\sigma $ errors. Also, the strategy
utilized with these datasets are talked about in details below.

\subsection{\textbf{Hz datasets}}

The Hubble parameter $H$ is one of the most crucial cosmic parameter that
explains the expansion rate in the Universe. It contains the necessary
information on the cosmic history. The Hubble parameter in terms of some
physical quantities such as redshift, length and time can be expressed as, $%
H(z)=-\frac{1}{1+z}\frac{dz}{dt}$, where $dz$ is the outcome of
spectroscopic surveys and the term $dt$ gives the model-independent value of 
$H(z)$. The observations from the parameter $H(z)$ highlight the dark ages
of the Universe such as issues based on dark matter and dark energy .

There are two elemental arrangements which are greatly used in literature to
estimate the value of $H(z)$ at any instant $z$:\newline
(1) The extraction of $H(z)$ from BAO data\newline
(2) The differential age methodology \cite{H1}-\cite{H19}.\newline
\qquad In this article, we have taken the updated list of datasets to $57$
points (i.e. $31$ and $26$ one from the differential age approach and other
from the BAO and some different strategies respectively) between the range
of redshift $z\in \lbrack 0.07,2.42]$ \cite{sharov}. Although there is
discrepancy in choosing the value of Hubble parameter (known as Hubble
tension), here we have taken $H_{0}=69$ $Km/s/Mpc$ for our analysis.

The observational constraints on the model parameters $\alpha $ and $\beta $
(proportionate to the highest probability examination) can be captured by
minimizing the chi-square value ($\chi _{min}^{2}$), i.e. is equivalent to
say the maximum likelihood analysis. The likelihood function $\chi
_{H}^{2}(\alpha ,\beta ,\Omega _{m0},\Omega _{\phi 0})$ can be computed as:

\begin{equation*}
\chi _{H}^{2}(\alpha ,\beta ,\Omega _{m0},\Omega _{\phi
0})=\sum\limits_{i=1}^{57}\frac{[H_{th}(z_{i},\alpha ,\beta ,\Omega
_{m0},\Omega _{\phi 0})-H_{obs}(z_{i})]^{2}}{\sigma _{H(z_{i})}^{2}},
\end{equation*}%
where $H_{obs}$ and $H_{th}$ assume the role of the observed and theoretical
value of $H$ respectively. Also, $\sigma _{H(z_{i})}$ indicate the standard
error in the value of $H$ so observed. Table-1 expresses $57$ points of $%
H(z) $ with corresponding errors $\sigma _{H}$ along with the references.

\subsection{\textbf{Pantheon datasets}}

As we move into the advancement of technology, we study observational
cosmology through various datasets which discuss various certainties in
observational cosmology (e.g., early evolution, structure formation, and
secrets of the dark Universe that somehow explain the late-time cosmic
acceleration employing cosmic mechanism and ray detectors). Among the
various observational datasets, the pantheon sample is one of the most
significant datasets, which contains $1048$ data points, and is the foremost
discharged supernovae type Ia datasets.

Pantheon datasets points \cite{SLONICPANTHEON} of spectroscopically wrap the
range of redshift $z$ in the interval $z\in (0.01,2.26)$. The outcome of
these datasets provide the assessment of the distance moduli $\mu _{i}=\mu
_{i}^{obs}$ in the range of redshift $z\in (0,1.41]$

The distance moduli can be obtained by the equation $\mu _{i}^{th}=\mu
(D_{L})=m-M=5\log _{10}(D_{L})+\mu _{0}$. The associated terms $M$, $m$ and $%
\mu _{0}$ in this equation are represented as absolute magnitude, apparent
magnitude and marginalized nuisance parameter respectively. Also, $\mu _{0}$
can be calculated as $\mu _{0}=5\log \left( H_{0}^{-1}/Mpc\right) +25$. To
identify the best match of the model parameters $\alpha $, $\beta $ and
cosmological parameters $\Omega _{m0}$ \& $\Omega _{\phi 0}$ of the
developed model, the theoretical value of the distance modulus ($\mu
_{i}^{th}$) can be compared with the observed value of distance modulus ($%
\mu _{i}^{obs}$). By the above data, the formula of luminosity distance can
be adjusted as:

\begin{eqnarray*}
D_{l}(z) &=&\frac{c(1+z)}{H_{0}}S_{k}\left( H_{0}\int_{0}^{z}\frac{1}{
H(z^{\ast })}dz^{\ast }\right) , \\
\text{where }S_{k}(x) &=&\left\{ 
\begin{array}{c}
\sinh (x\sqrt{\Omega _{k}})/\Omega _{k},\Omega _{k}>0 \\ 
x,\ \ \ \ \ \ \ \ \ \ \ \ \ \ \ \ \ \ \ \ \ \ \ \Omega _{k}=0 \\ 
\sin x\sqrt{\left\vert \Omega _{k}\right\vert })/\left\vert \Omega
_{k}\right\vert \ ,\Omega _{k}<0%
\end{array}
\right. .
\end{eqnarray*}%
\qquad

Here, we have the density parameter of flat space-time is $\Omega _{k}=0$.
Moreover, to compute the variation between the SN Ia data and the
predictions of our model, we need the luminosity distance $D_{l}(z)$ and the
chi square function. Pantheon datasets requires the $\chi _{SN}^{2}$
function in the form of,

\begin{equation}
\chi _{SN}^{2}(\alpha ,\beta ,\Omega _{m0},\Omega _{\phi
0})=\sum\limits_{i=1}^{1048}\frac{[\mu ^{th}(\alpha ,\beta ,\Omega
_{m0},\Omega _{\phi 0})-\mu ^{obs}(z_{i})]^{2}}{\sigma _{\mu (z_{i})}^{2}},
\label{chisn}
\end{equation}%
where $\sigma _{\mu (z_{i})}^{2}$ stands for the standard error of the
observed value.

\subsection{\textbf{BAO datasets}}

The study of baryonic acoustic oscillations takes place in the early
universe, when baryons and photons are inextricably linked thanks to Thomson
scattering. Due to the high pressure of photons, both baryons and photons
operate as a single fluid that cannot collapse under gravity but rather
oscillates and gives the name to these oscillations as Baryonic acoustic
oscillations (BAO). The $BAO$ entails calculating the structural
distribution of galaxies to control the rate at which cosmic structure grows
throughout the universe's overall expansion. This difference, in theory, can
discriminate between different forms of DE. Galaxy clustering patterns are
based on statistics that describe how little discrepancies in cosmic
structure are magnified. This clustering encodes a significant average
parting among galaxies, which might be used to reconstruct the universe's
expansion history in the same way as supernovae type Ia (standard candles )
are used. The sound horizon $r_{s}$, when a photon decouples at the redshift 
$z_{\ast }$, regulates the characteristic scale of BAO, which is delivered
as,

\begin{equation*}
r_{s}(z_{\ast })=\frac{c}{\sqrt{3}}\int_{0}^{\frac{1}{1+z_{\ast }}}\frac{da}{
a^{2}H(a)\sqrt{1+(3\Omega _{0b}/4\Omega _{0\gamma })a}}.
\end{equation*}

Here, the terms $\Omega _{0b}$ and $\Omega _{0\gamma }$ mean the physical
quantities demonstrated as baryon density and photon density at present time 
$t$.

The angular diameter distance $D_{A}$ and the Hubble function $H(z)$ can
also be extracted using $r_{s}$ (sound horizon scale of BAO). Let $\triangle
\theta $ and $\triangle z$ are the observed angular separation and the
measured redshift separation of the BAO feature in the 2 point correlation
function of the galaxy distribution on the sky is $\triangle \theta $, and
of the BAO feature in the 2 point correlation function along the line of
sight is $\triangle z$, then $\triangle \theta =\frac{r_{s}}{d_{A}(z)}$
where $d_{A}(z)=\int_{0}^{z}\frac{dz^{\prime }}{H(z^{\prime })}$ and $%
\triangle z=H(z)r_{s}$.\newline

BAO datasets of $d_{A}(z)/DV(z_{BAO})$ are studied in the literature \cite%
{BAO1}, \cite{BAO2}, \cite{BAO3}, \cite{BAO4}, \cite{BAO5}, \cite{BAO6}
where the photon decouples at redshift $z_{\ast }\approx 1091$, $d_{A}(z)$
represents the co-moving angular diameter distance, and $D_{V}(z)=\left(
d_{A}(z)^{2}z/H(z)\right) ^{1/3}$ highlights the dilation scale. Table 2
shows the data that was used in this investigation.

\begin{center}
\begin{tabular}{|c|c|c|c|c|c|c|}
\hline
\multicolumn{7}{|c|}{Table-2: Values of $d_{A}(z_{\ast })/D_{V}(z_{BAO})$
for distinct values of $z_{BAO}$} \\ \hline
$z_{BAO}$ & $0.106$ & $0.2$ & $0.35$ & $0.44$ & $0.6$ & $0.73$ \\ \hline
$\frac{d_{A}(z_{\ast })}{D_{V}(z_{BAO})}$ & $30.95\pm 1.46$ & $17.55\pm 0.60$
& $10.11\pm 0.37$ & $8.44\pm 0.67$ & $6.69\pm 0.33$ & $5.45\pm 0.31$ \\ 
\hline
\end{tabular}
\end{center}

\qquad The chi square function for BAO is given by \cite{BAO6} 
\begin{equation}
\chi _{BAO}^{2}=X^{T}C^{-1}X\,,  \label{chibao}
\end{equation}
where 
\begin{equation}
X=\left( 
\begin{array}{c}
\frac{d_{A}(z_{\star })}{D_{V}(0.106)}-30.95 \\ 
\frac{d_{A}(z_{\star })}{D_{V}(0.2)}-17.55 \\ 
\frac{d_{A}(z_{\star })}{D_{V}(0.35)}-10.11 \\ 
\frac{d_{A}(z_{\star })}{D_{V}(0.44)}-8.44 \\ 
\frac{d_{A}(z_{\star })}{D_{V}(0.6)}-6.69 \\ 
\frac{d_{A}(z_{\star })}{D_{V}(0.73)}-5.45%
\end{array}
\right) \,,
\end{equation}
and $C^{-1}$ is the inverse covariance matrix defined in \cite{BAO6}.

\begin{equation}
C^{-1}=\left( 
\begin{array}{cccccc}
0.48435 & -0.101383 & -0.164945 & -0.0305703 & -0.097874 & -0.106738 \\ 
-0.101383 & 3.2882 & -2.45497 & -0.0787898 & -0.252254 & -0.2751 \\ 
-0.164945 & -2.454987 & 9.55916 & -0.128187 & -0.410404 & -0.447574 \\ 
-0.0305703 & -0.0787898 & -0.128187 & 2.78728 & -2.75632 & 1.16437 \\ 
-0.097874 & -0.252254 & -0.410404 & -2.75632 & 14.9245 & -7.32441 \\ 
-0.106738 & -0.2751 & -0.447574 & 1.16437 & -7.32441 & 14.5022%
\end{array}
\right) \,
\end{equation}

\begin{figure}[tbp]
\centering
\includegraphics[width=0.7\linewidth]{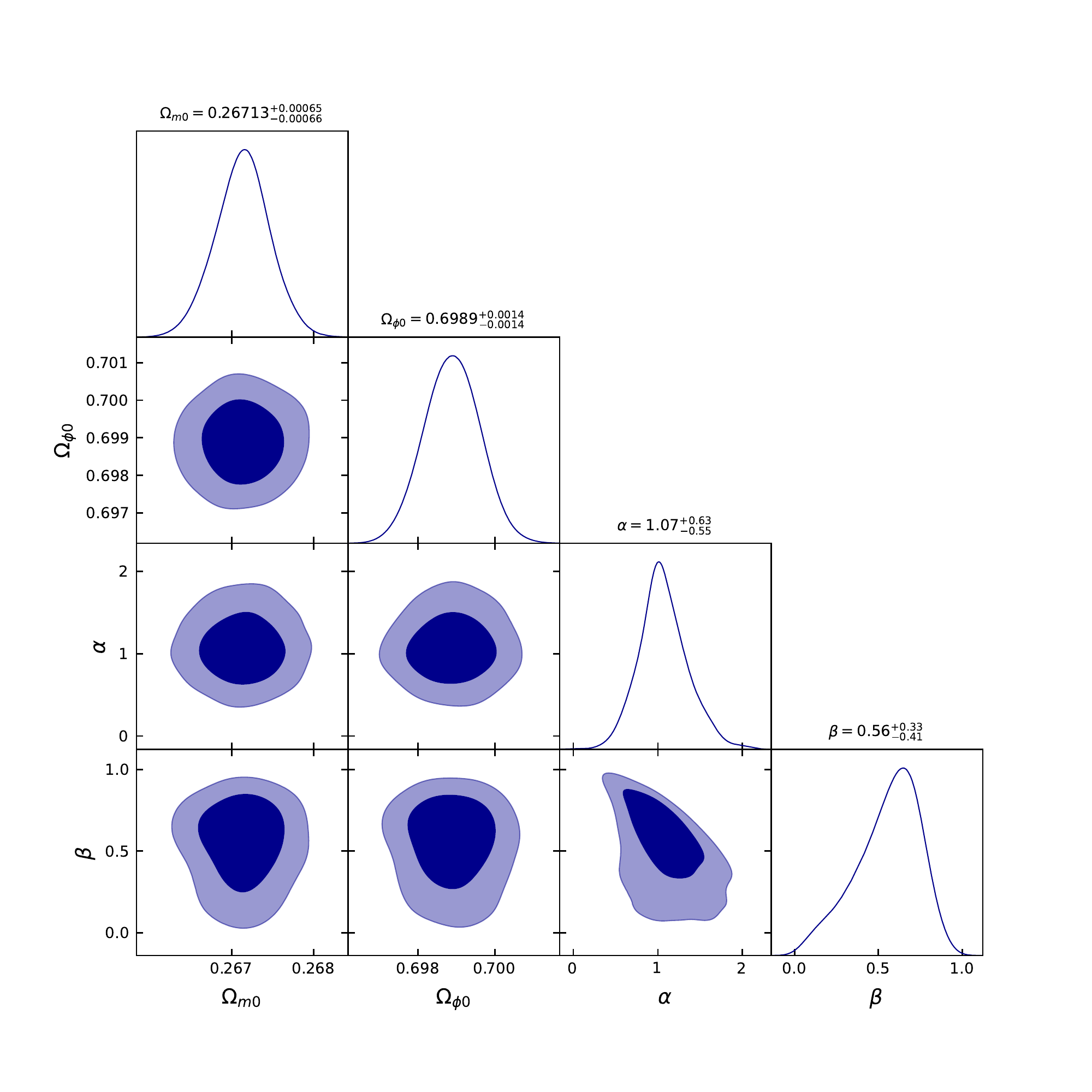}
\caption{{\protect\scriptsize {}The plot shows the two dimensional contour
plot of the four model parameters }$\protect\alpha ${\protect\scriptsize , }$%
\protect\beta ${\protect\scriptsize , }$\Omega _{m0}${\protect\scriptsize , }
$\Omega _{\protect\phi 0}${\protect\scriptsize \ of our model with }$1- 
\protect\sigma ${\protect\scriptsize \ and }$2-\protect\sigma $ 
{\protect\scriptsize \ errors and shows the best fit values of }$\protect%
\alpha ${\protect\scriptsize , }$\protect\beta ${\protect\scriptsize , }$%
\Omega _{m0}${\protect\scriptsize , }$\Omega _{\protect\phi 0}$ 
{\protect\scriptsize \ with respect to the }$57${\protect\scriptsize \
points of Hubble datasets as compiled in \protect\cite{sharov}.}}
\label{hubble}
\end{figure}

\begin{figure}[tbp]
\centering
\includegraphics[width=0.7\linewidth]{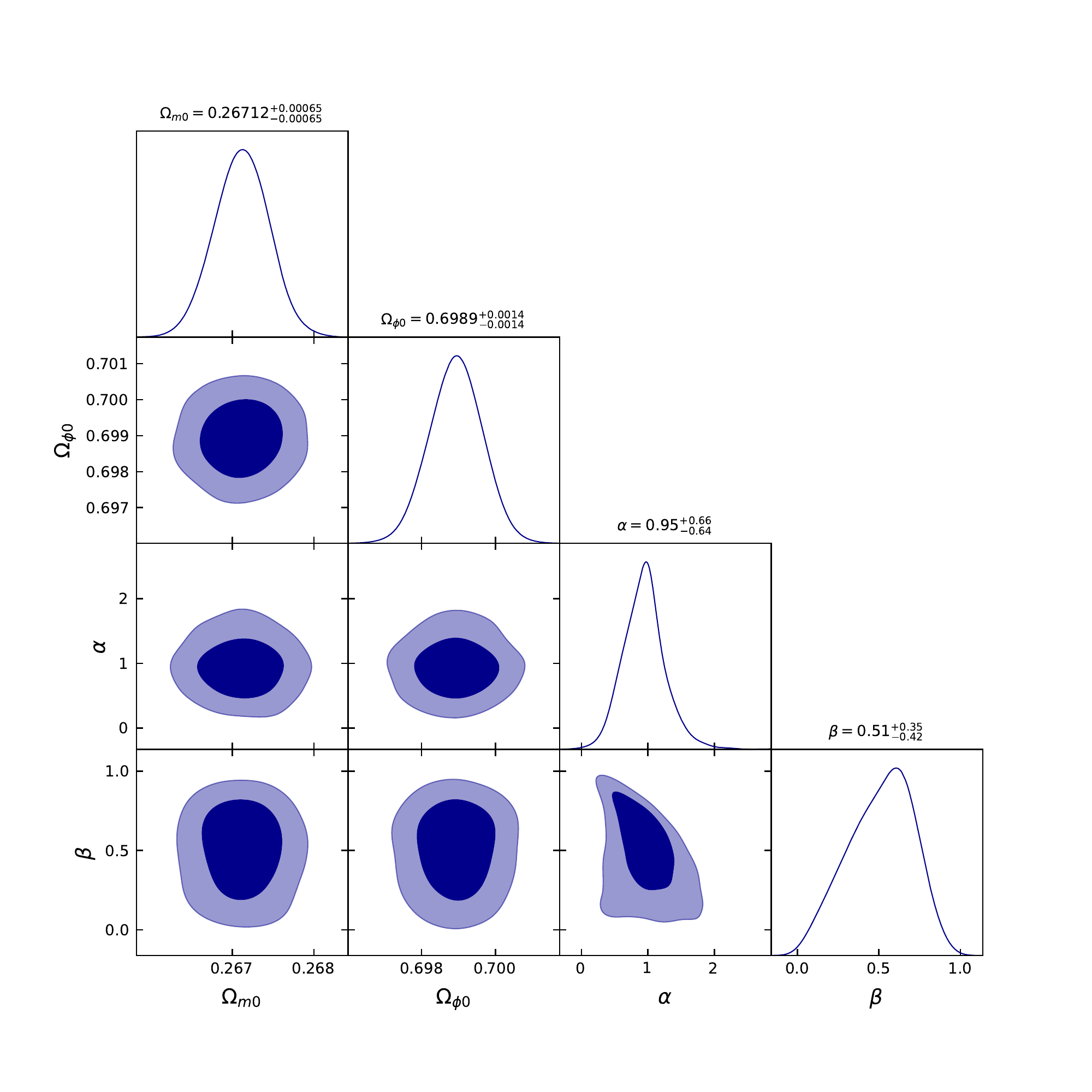}
\caption{{\protect\scriptsize {}The plot shows the two dimensional contour
plot of the four model parameters }$\protect\alpha ${\protect\scriptsize , }$%
\protect\beta ${\protect\scriptsize , }$\Omega _{m0}${\protect\scriptsize , }%
$\Omega _{\protect\phi 0}${\protect\scriptsize \ of our model with }$1-%
\protect\sigma ${\protect\scriptsize \ and }$2-\protect\sigma $%
{\protect\scriptsize \ errors and shows the best fit values of }$\protect%
\alpha ${\protect\scriptsize , }$\protect\beta ${\protect\scriptsize , }$%
\Omega _{m0}${\protect\scriptsize , }$\Omega _{\protect\phi 0}$%
{\protect\scriptsize \ with respect to the }$1048${\protect\scriptsize \
points of Pantheon datasets as compiled in \protect\cite{SLONICPANTHEON}.{}}}
\label{pantheon}
\end{figure}

\begin{figure}[tbp]
\centering
\includegraphics[width=0.7\linewidth]{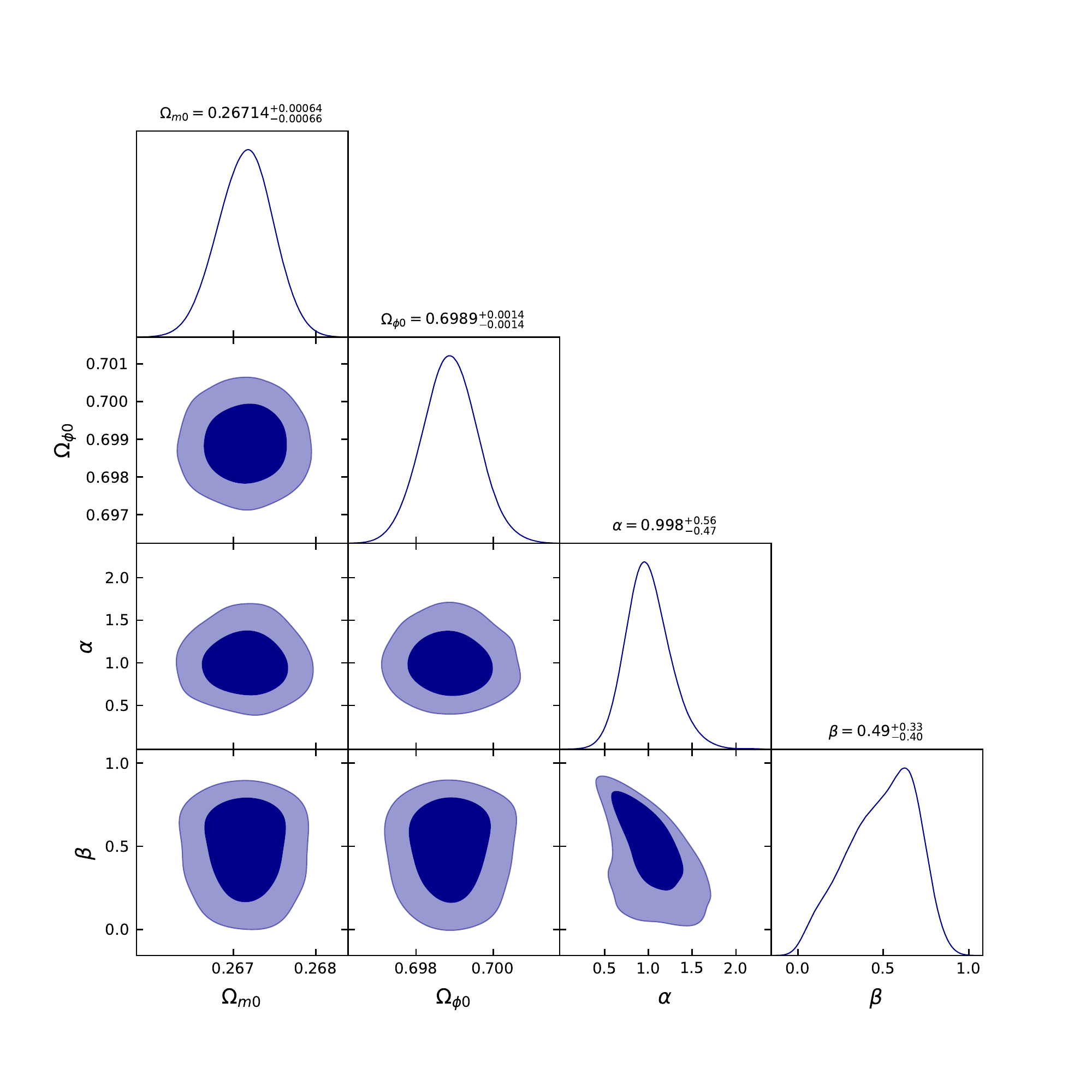}
\caption{{\protect\scriptsize {}{}This plot shows the }$2-D$%
{\protect\scriptsize \ contour plot of the four model parameters }$\protect%
\alpha ${\protect\scriptsize , }$\protect\beta ${\protect\scriptsize , }$%
\Omega _{m0}${\protect\scriptsize , }$\Omega _{\protect\phi 0}$%
{\protect\scriptsize \ of our model with }$1-\protect\sigma $%
{\protect\scriptsize \ and }$2-\protect\sigma ${\protect\scriptsize \ errors
and shows the best fit values of }$\protect\alpha ${\protect\scriptsize , }$%
\protect\beta ${\protect\scriptsize , }$\Omega _{m0}${\protect\scriptsize , }%
$\Omega _{\protect\phi 0}${\protect\scriptsize \ with respect to the
combined Hubble \protect\cite{sharov}, Pantheon \protect\cite{SLONICPANTHEON}
and BAO \protect\cite{BAO6} datasets.}}
\label{hubpanthbao}
\end{figure}

\begin{figure}[tbp]
\centering
\includegraphics[width=5.5 in, height=3 in]{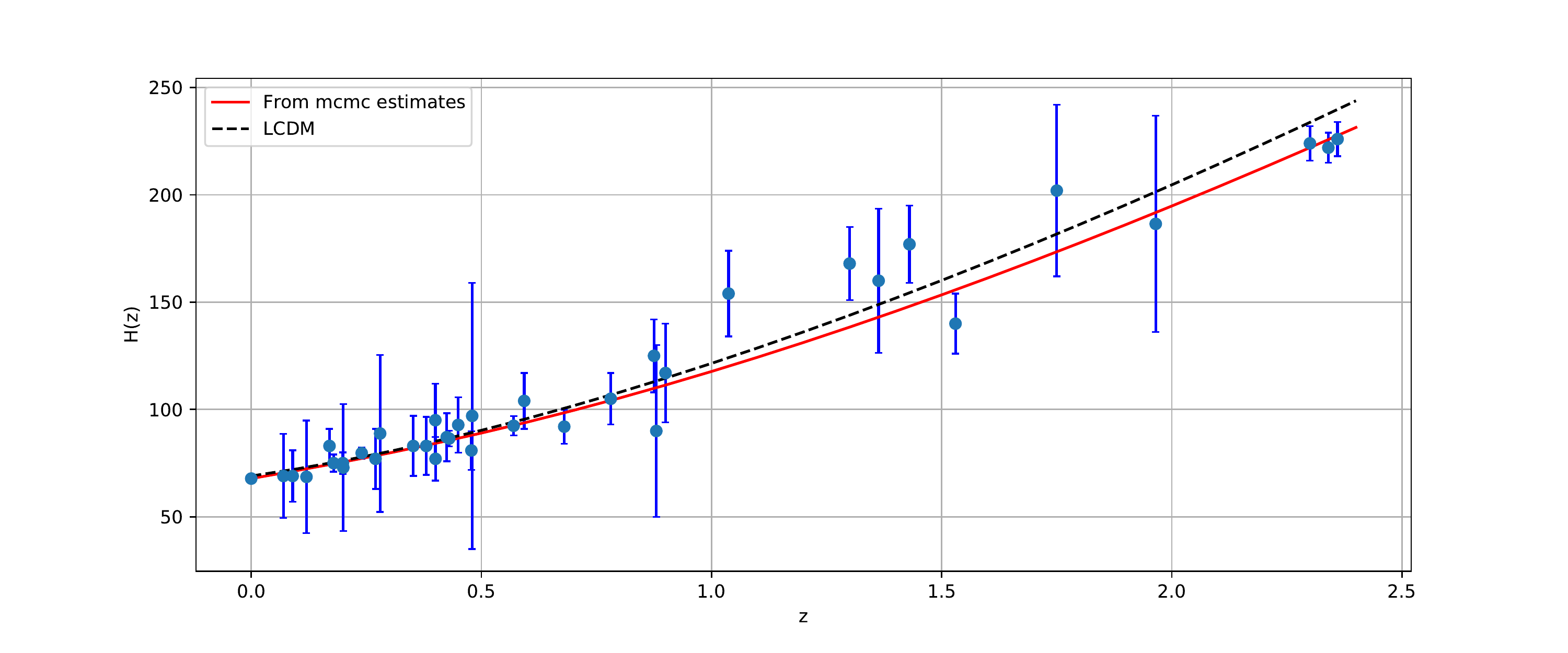}
\caption{{\protect\scriptsize {}{}The plot of Hubble function }$H(z)$%
{\protect\scriptsize \ vs. redshift }$z${\protect\scriptsize \ for our
obtained model with the best fit values of the model parameters and compared
with the }$\Lambda ${\protect\scriptsize CDM model showing nice fit to the }$%
57${\protect\scriptsize \ points of the considered Hubble datasets with
errorbars.}}
\label{h-z}
\end{figure}

\begin{figure}[tbp]
\centering
\includegraphics[width=5.5 in, height=3 in]{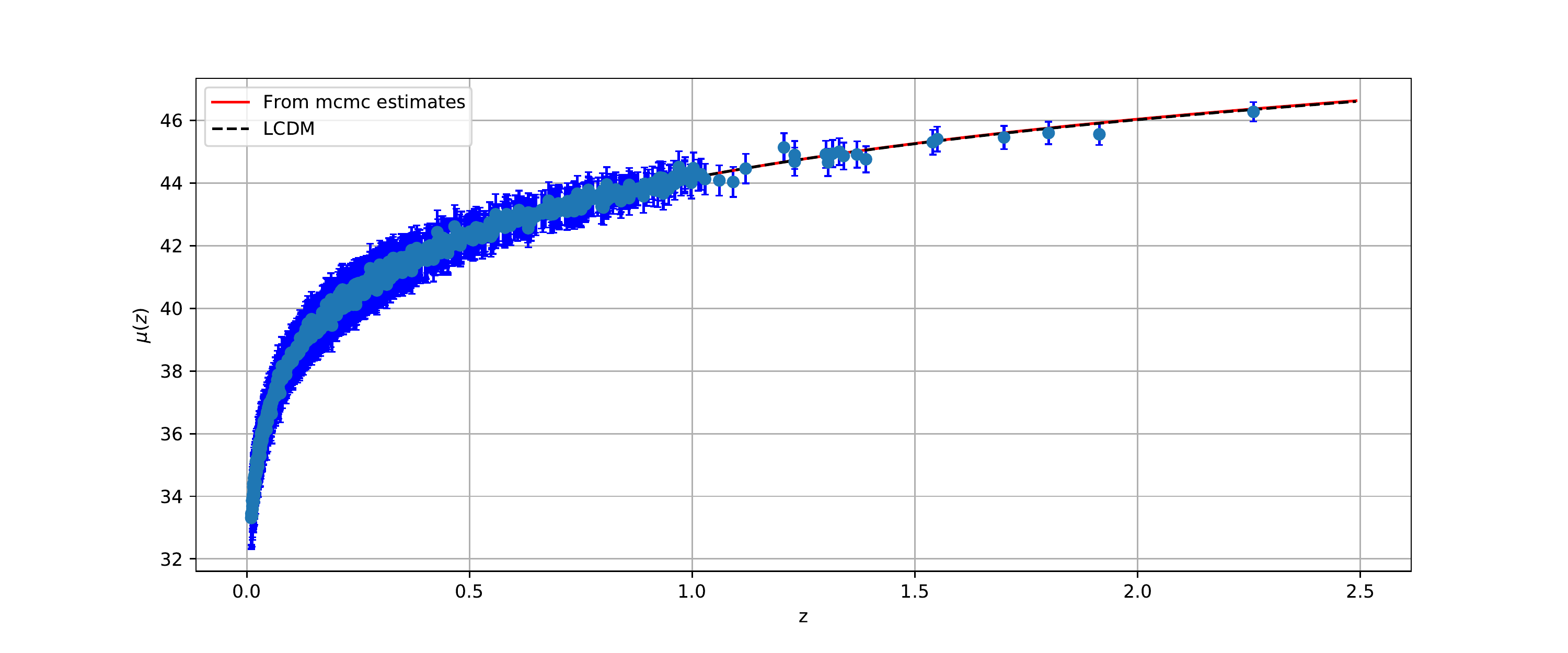}
\caption{{\protect\scriptsize {}{}{}The plot of }$\protect\mu (z)$%
{\protect\scriptsize \ vs. }$z${\protect\scriptsize \ for our obtained model
with the best fit values of the model parameters and compared with the }$%
\Lambda ${\protect\scriptsize CDM model showing nice fit to the }$1048$%
{\protect\scriptsize \ points of the considered Pantheon datasets with
errorbars.}}
\label{mu-z}
\end{figure}

By minimising the chi-square, we were able to get the best fit values of the
model parameters $\alpha $, $\beta $, and cosmological parameters $\Omega
_{m0}$ \& $\Omega _{\phi 0}$ as two-dimensional contour plots with $1-\sigma 
$\ and $2-\sigma $ errors using the datasets presented above. The following
figures show the contour plots for $\alpha $, $\beta ,\Omega _{m0},\Omega
_{\phi 0}$\ with respect to the Hubble datasets Fig. \ref{hubble}, Pantheon
datasets Fig. \ref{pantheon} and combined Hubble+Pantheon+BAO datasets Fig. %
\ref{hubpanthbao}. The model parameter values with the best fit are
determined as follows: $\alpha =1.07_{-0.55}^{+0.63}$, $\beta
=0.56_{-0.41}^{+0.33}$, $\Omega _{m0}=0.26713_{-0.00066}^{+0.00065}$, $%
\Omega _{\phi 0}=0.6989_{-0.0014}^{+0.0014}$\ for Hubble datasets, $\alpha
=0.95_{-0.54}^{+0.66}$, $\beta =0.51_{-0.42}^{+0.35}$, $\Omega
_{m0}=0.26712_{-0.00065}^{+0.00065}$, $\Omega _{\phi
0}=0.6989_{-0.0014}^{+0.0014}$\ for Pantheon datasets and $\alpha
=0.998_{-0.47}^{+0.56}$, $\beta =0.49_{-0.40}^{+0.33}$, $\Omega
_{m0}=0.26714_{-0.00066}^{+0.00064}$, $\Omega _{\phi
0}=0.6989_{-0.0014}^{+0.0014}$ for combined Hubble+Pantheon+BAO datasets.
The error bar charts for the aforementioned Hubble datasets are also
displayed in Fig. \ref{h-z} and Pantheon datasets in the Fig. \ref{mu-z} and
compared our obtained model (Red line) with the $\Lambda $CDM model (with $%
\Omega _{\Lambda 0}=0.7$\ and $\Omega _{m0}=0.3$) shown in dashed line. Our
model shows nice fit to both the datasets.

We now have all of the theoretical formulae as well as numerical values for
the model parameters, and we can examine the model's physical dynamics. As a
result, the physical dynamics of the other essential cosmological parameters
will be discussed in the next section.

\section{Physical Dynamics of the Model}

\label{sec6}

The physical dynamics of the obtained model can be explained through the
behavior of physical and geometrical parameters. So, in this section, we
will discuss the behavior of the important cosmological parameters in the
late times and future.

\subsection{Deceleration parameter}

The deceleration parameter $q<0$ indicates that the Universe is expanding at
a faster pace, whereas $q>0$ indicates that it is slowing down. The
decelerating phase is important in the cosmic evolution for the structure
formation in the Universe, and the accelerating phase in the late Universe
can explain the SNe Ia data. That means there must be a phase of transition (%
$q=0$). In our model, the expression for the deceleration parameter $q=-%
\frac{\ddot{a}a}{\dot{a}^{2}}=-1+\left( \frac{1+z}{H}\right) \frac{dH}{dz}$
in terms of redshift $z$, involving the model parameters $\alpha $, $\beta $%
, $\Omega _{m0}$, $\Omega _{\phi 0}$ is given by,

\begin{eqnarray}
q(z) &=&-1+  \notag \\
&&\frac{(1+z)\left\{ 3\text{$\Omega _{m0}$}(1+z)^{2}+\text{$\Omega _{\phi 0}$%
}(1+z)^{-\beta }sech^{-1}\left( \beta \right) ^{-\alpha }\cosh ^{-1}\left( 
\frac{1+z}{\beta }\right) ^{\alpha }\left( \frac{\alpha }{\sqrt{\frac{%
1+z-\beta }{1+z+\beta }}(1+z+\beta )\cosh ^{-1}\left( \frac{1+z}{\beta }%
\right) }-\frac{\beta }{1+z}\right) \right\} }{2\left\{ \text{$\Omega _{m0}$}%
(1+z)^{3}+\text{$\Omega _{\phi 0}$}(1+z)^{-\beta }sech^{-1}(\beta )^{-\alpha
}\cosh ^{-1}\left( \frac{1+z}{\beta }\right) ^{\alpha }\right\} }  \label{qz}
\end{eqnarray}

The graphic below depicts the evolution of the deceleration parameter, which
describes the late evolution of a phase transition including early slowdown
and late acceleration.

\begin{figure}[tbp]
\begin{center}
$%
\begin{array}{c@{\hspace{0.1in}}c}
\includegraphics[width=3.0 in, height=2.5 in]{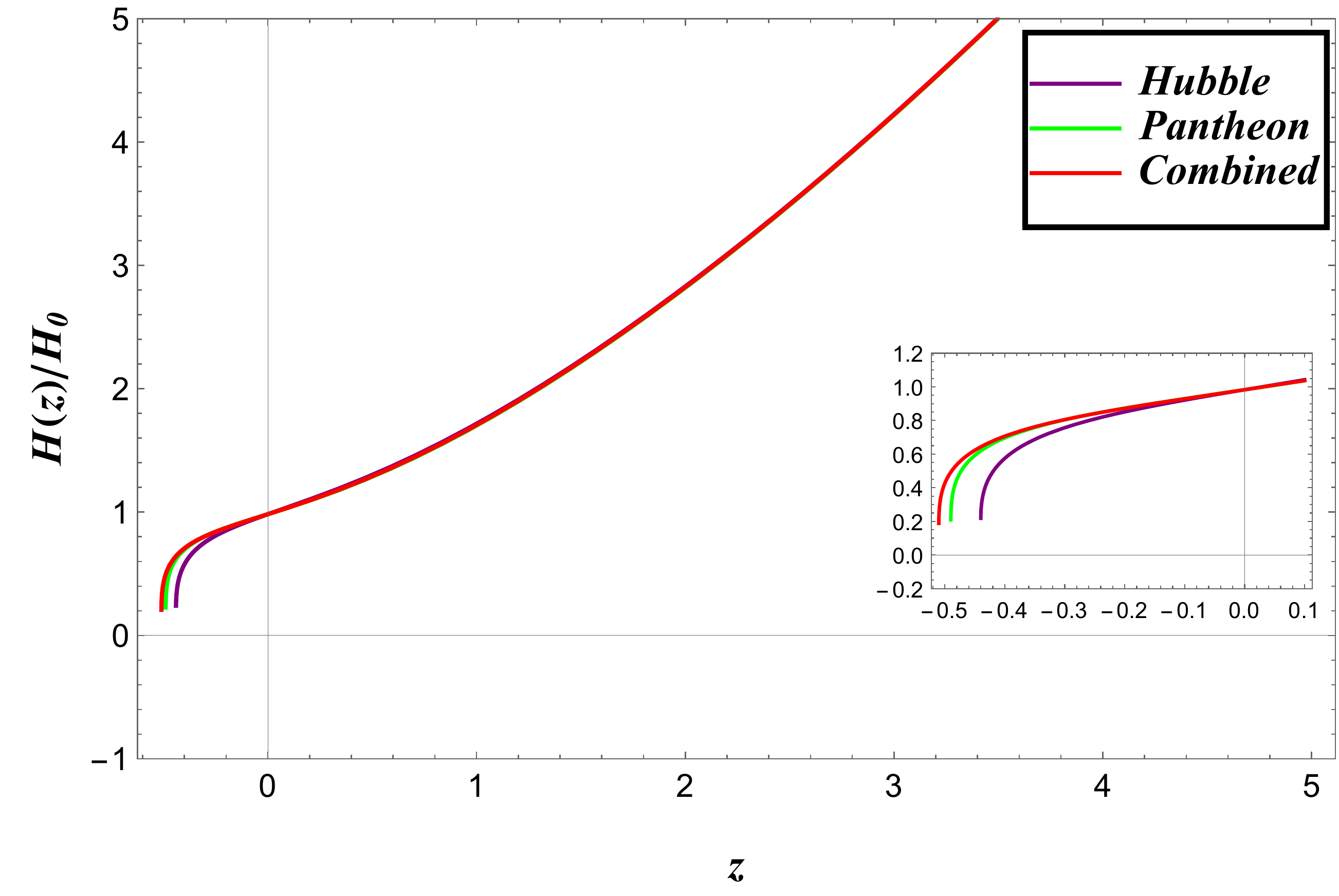} & %
\includegraphics[width=3.0 in, height=2.5 in]{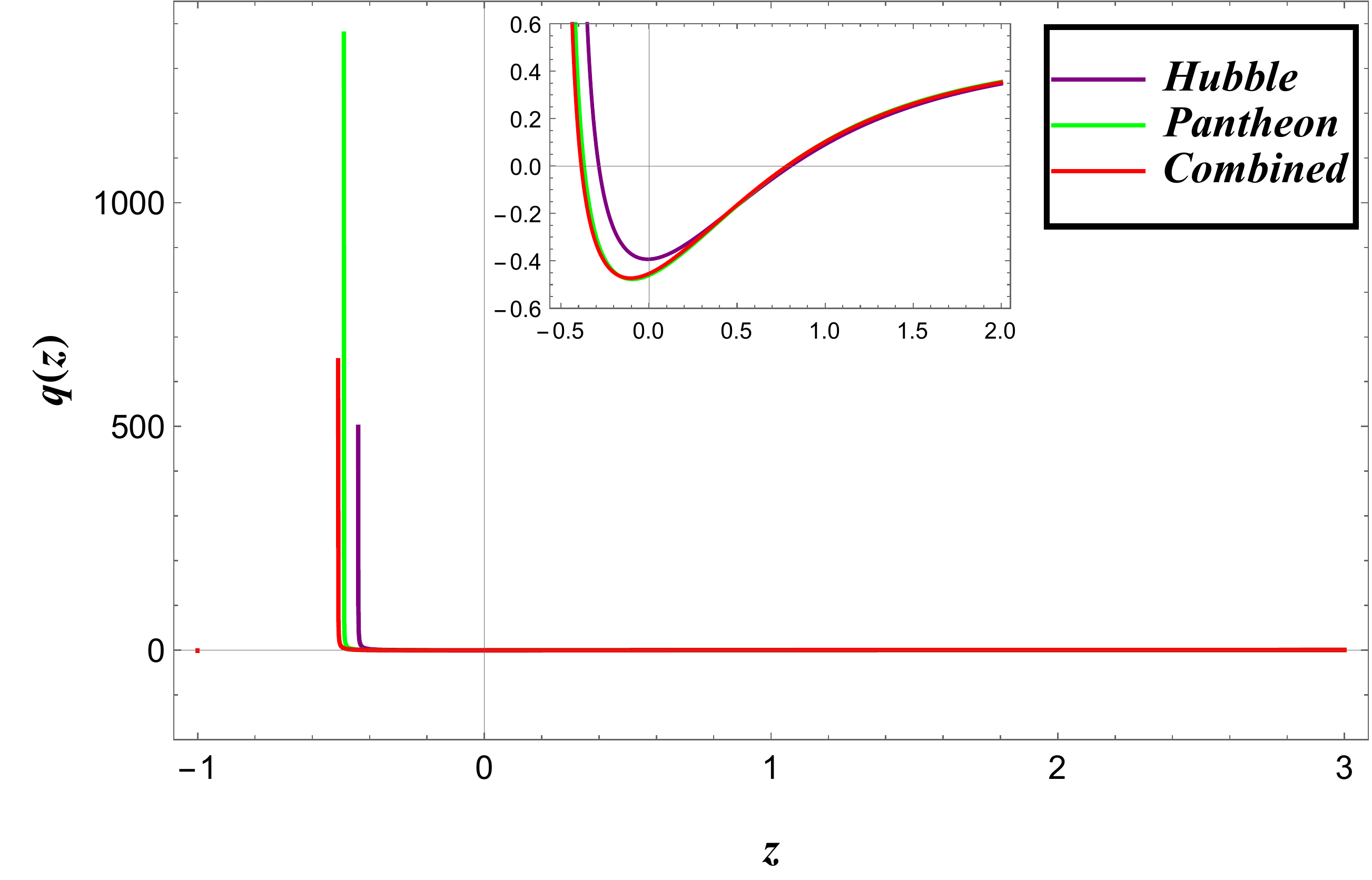} \\ 
\mbox (a) & \mbox (b)%
\end{array}
$%
\end{center}
\caption{{\protect\scriptsize In the figure the left panel (a) shows the
evolution of the Hubble parameter }$H${\protect\scriptsize \ \ \ vs.
redshift }$z${\protect\scriptsize \ and the right panel (b) shows the
evolution of the deceleration parameter }$q${\protect\scriptsize \ \ \ vs.
redshift }$z${\protect\scriptsize \ for the different sets of best fitted
values of the model parameters }$\protect\alpha ${\protect\scriptsize , }$%
\protect\beta ${\protect\scriptsize , }$\Omega _{m0}${\protect\scriptsize , }%
$\Omega _{\protect\phi 0}${\protect\scriptsize \ obtained from different
observational datasets.}}
\label{H-z & q-z}
\end{figure}

In the Fig. \ref{H-z & q-z}, the left panel (a) highlights the recent past
and future evolution Hubble parameter $H$. The future evolution is distinct
from that of standard scenario. Similarly, the right panel (b) shows the
evolution of the deceleration parameter $q$ indicating decelerating
expansion in the past with a phase transitions at $z\approx 0.79$. The
Universe then accelerates for an extended length of time owing to dark
energy dominance. The accelerating phase of expansion in our model
corresponds to the data, and this phase will continue until $z\approx -0.29$%
. From the plot of deceleration parameter, we can see a different behavior
of the Universe from the standard lore. The deceleration parameter becomes
zero at $z\approx -0.29$ and becomes positive afterwards which infers that
the Universe start to contract and collapse to a $Big\,Crunch$ singularity.

\subsection{Evolution of energy densities \& pressure}

\begin{figure}[tbp]
\begin{center}
$%
\begin{array}{c@{\hspace{.1in}}cc}
\includegraphics[width=2.2 in, height=2.2 in]{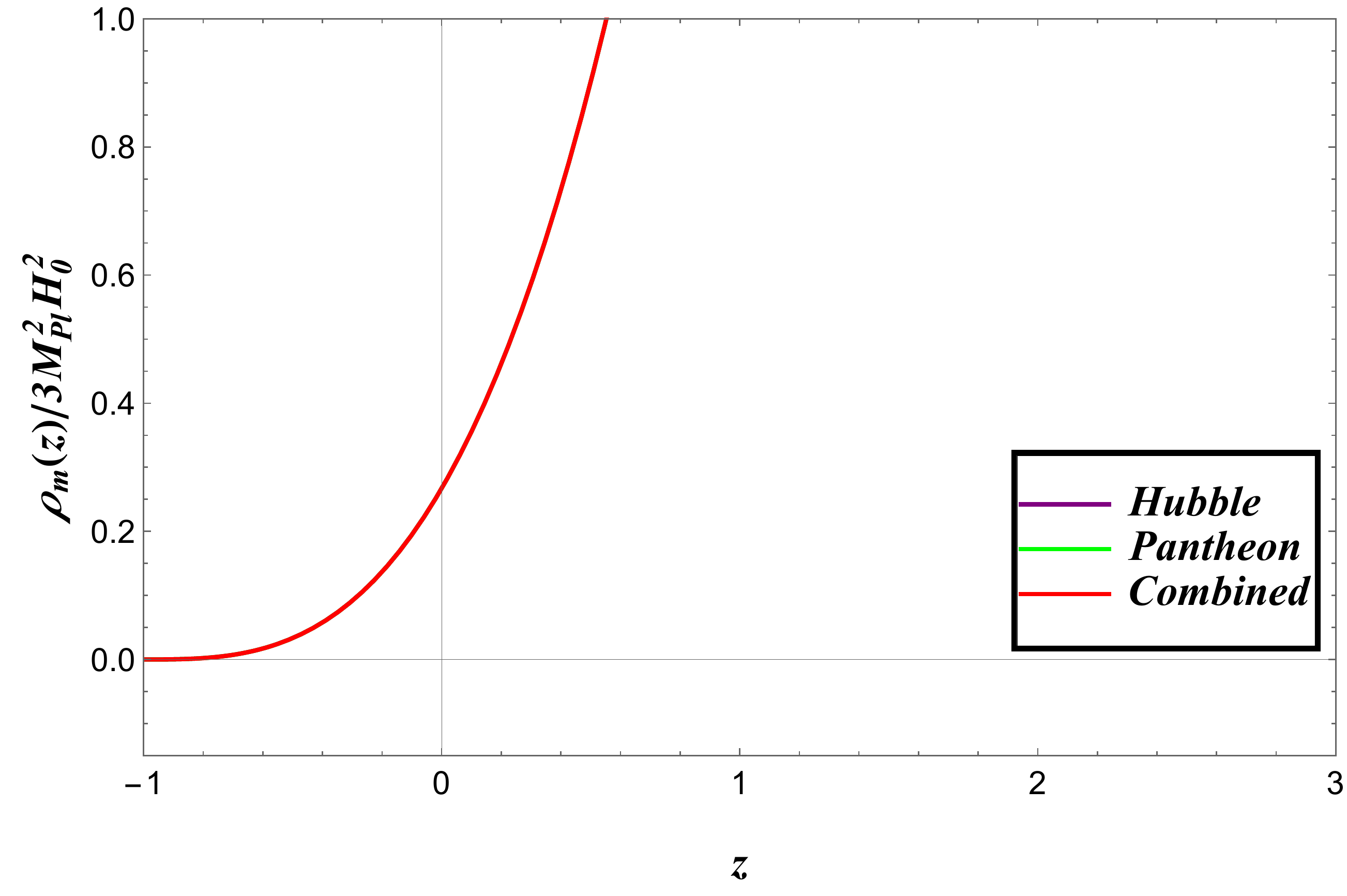} & %
\includegraphics[width=2.2 in, height=2.2 in]{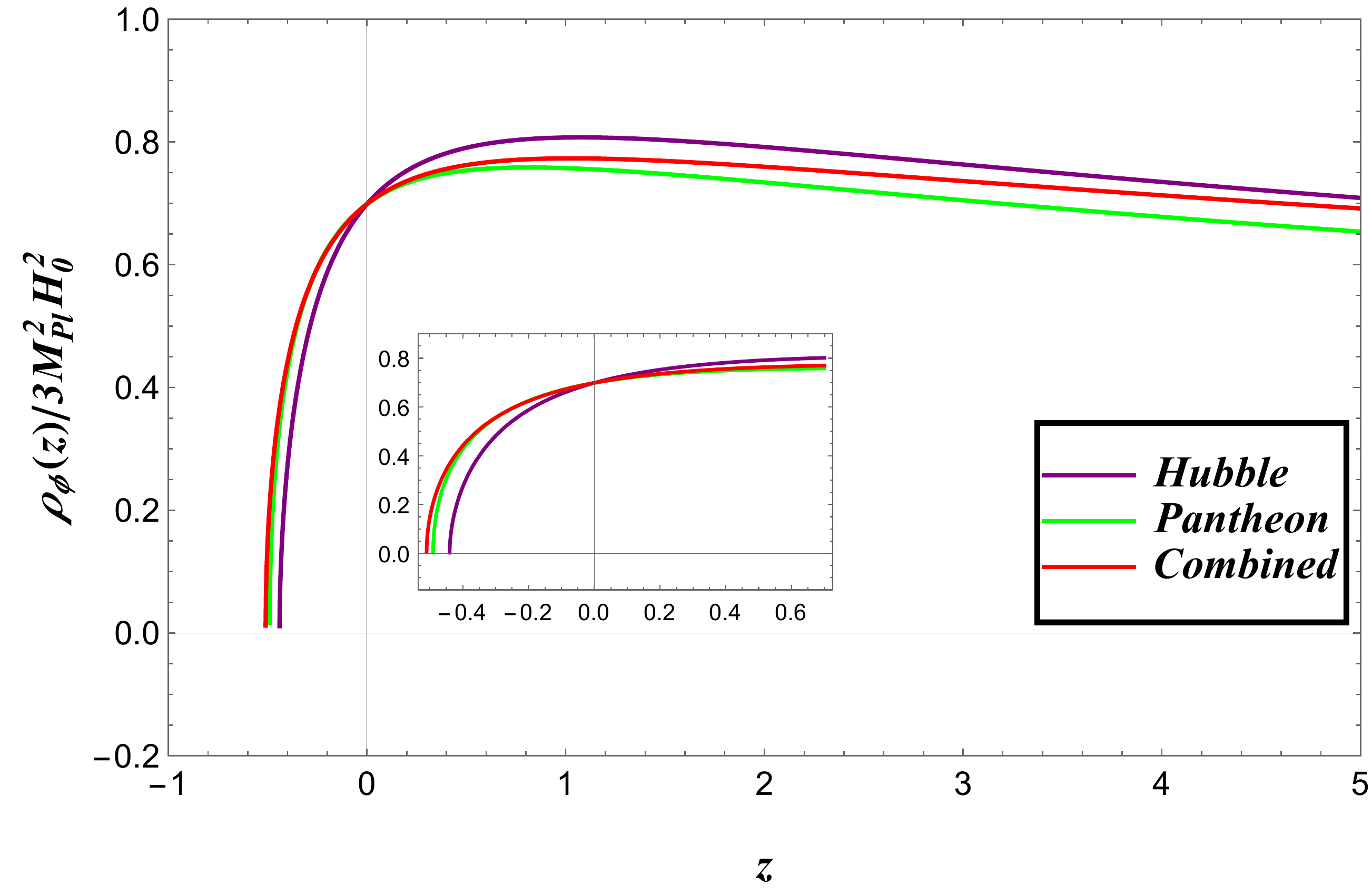} & %
\includegraphics[width=2.2 in, height=2.2 in]{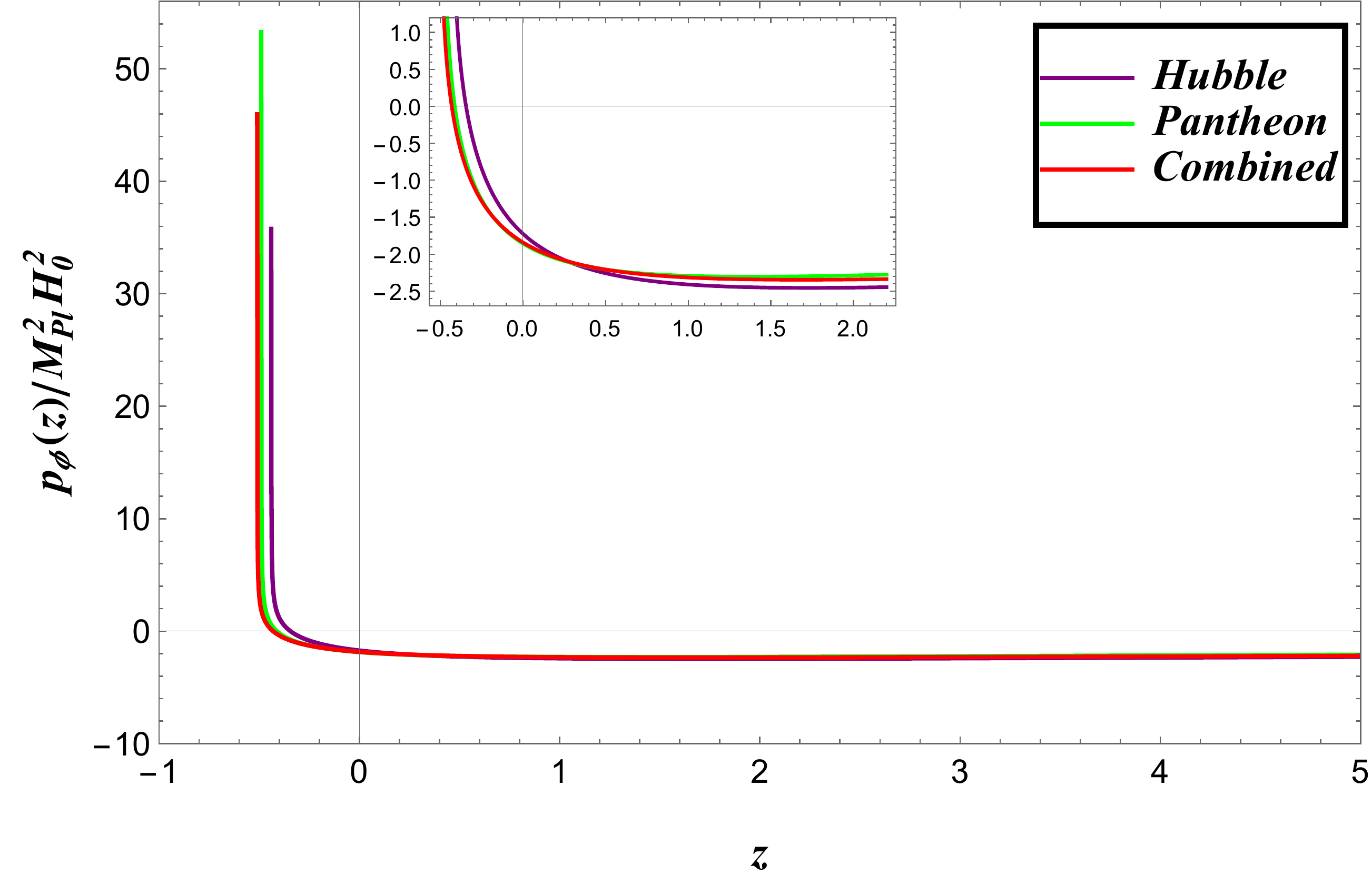} \\ 
\mbox (a) & \mbox (b) & \mbox (c)%
\end{array}
$%
\end{center}
\caption{{\protect\scriptsize The figure shows the plots of energy density
of matter }$\protect\rho _{m}${\protect\scriptsize , energy density of
scalar field }$\protect\rho _{\protect\phi }${\protect\scriptsize \ and
pressure }$p_{\protect\phi }${\protect\scriptsize \ for our obtained model
in three different panels. The three different lines shown in different
colors are our models for different sets of values of model parameters. }}
\label{densities}
\end{figure}

The evolution of matter and dark energy energy densities, as well as the
pressure of dark energy, is depicted in Fig. \ref{densities} based on the
values of model parameters derived from three distinct statistical studies.
The first panel (a) exhibits the role of the energy density of matter which
is initially very high and tends to zero in the near future. The second
panel (b) shows the behavior of the dark energy density $\rho _{\phi }$,
which is high in the past, decreases in a concave downward way as time
departed and finally tends to $0$ as expected from the above discussions of
Hubble and deceleration parameters. The third panel (c) highlights the
profile of pressure of dark energy (scalar field). Initially, $p_{\phi }$ is
negative, and assumes a highly positive value in the future showing a
contracting phase and collapse to a big crunch. \newline

\subsection{EoS parameter}

The expressions for the equation of state parameter $\omega _{Total}=\frac{%
p_{Total}}{\rho Total}=\frac{2q-1}{3}$ and $\omega _{\phi }=\frac{p_{\phi }}{%
\rho _{\phi }}=\frac{M_{pl}^{2}H^{2}\left( 2q-1\right) }{3M_{pl}^{2}H^{2}-%
\rho _{m}}$ are obtained in terms of redshift $z$, involving the model
parameters $\alpha $, $\beta $, $\Omega _{m0}$ and $\Omega _{\phi 0}$ are
given by, 
\begin{equation}
\omega _{\phi }(z)=\frac{\alpha (1+z)-(3+\beta )\sqrt{\frac{1+z-\beta }{%
1+z+\beta }}(1+z+\beta )\cosh ^{-1}\left( \frac{1+z}{\beta }\right) }{3\sqrt{%
\frac{1+z-\beta }{1+z+\beta }}(1+z+\beta )\cosh ^{-1}\left( \frac{1+z}{\beta 
}\right) }  \label{18rs}
\end{equation}

\begin{equation}
\omega _{Total}(z)=\frac{\text{$\Omega _{\text{$\phi 0$}}$}\cosh ^{-1}\left( 
\frac{1+z}{\beta }\right) ^{\alpha -1}\left( \alpha (1+z)-(3+\beta )\sqrt{%
\frac{1+z-\beta }{1+z+\beta }}(1+z+\beta )\cosh ^{-1}\left( \frac{1+z}{\beta 
}\right) \right) }{3\sqrt{\frac{1+z-\beta }{1+z+\beta }}(1+z+\beta )\left( 
\text{$\Omega _{m\text{$0$}}$}(1+z)^{3+\beta }sech^{-1}(\beta )^{\alpha }+%
\text{$\Omega _{\text{$\phi 0$}}$}\cosh ^{-1}\left( \frac{1+z}{\beta }%
\right) ^{\alpha }\right) }  \label{18rs1}
\end{equation}%
\begin{figure}[tbp]
\begin{center}
$%
\begin{array}{c@{\hspace{0.1in}}c}
\includegraphics[width=3.0 in, height=2.5 in]{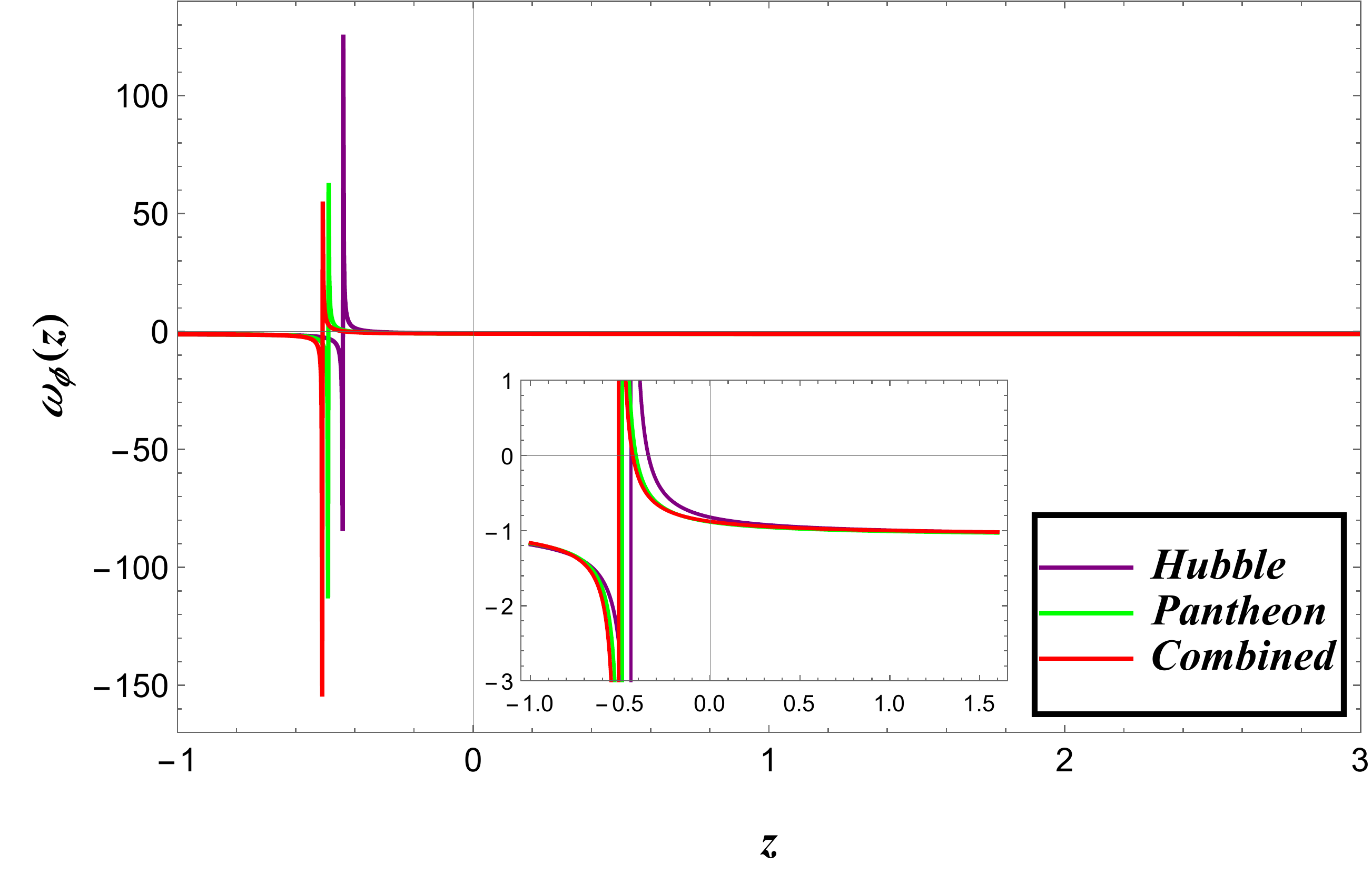} & %
\includegraphics[width=3.0 in, height=2.5 in]{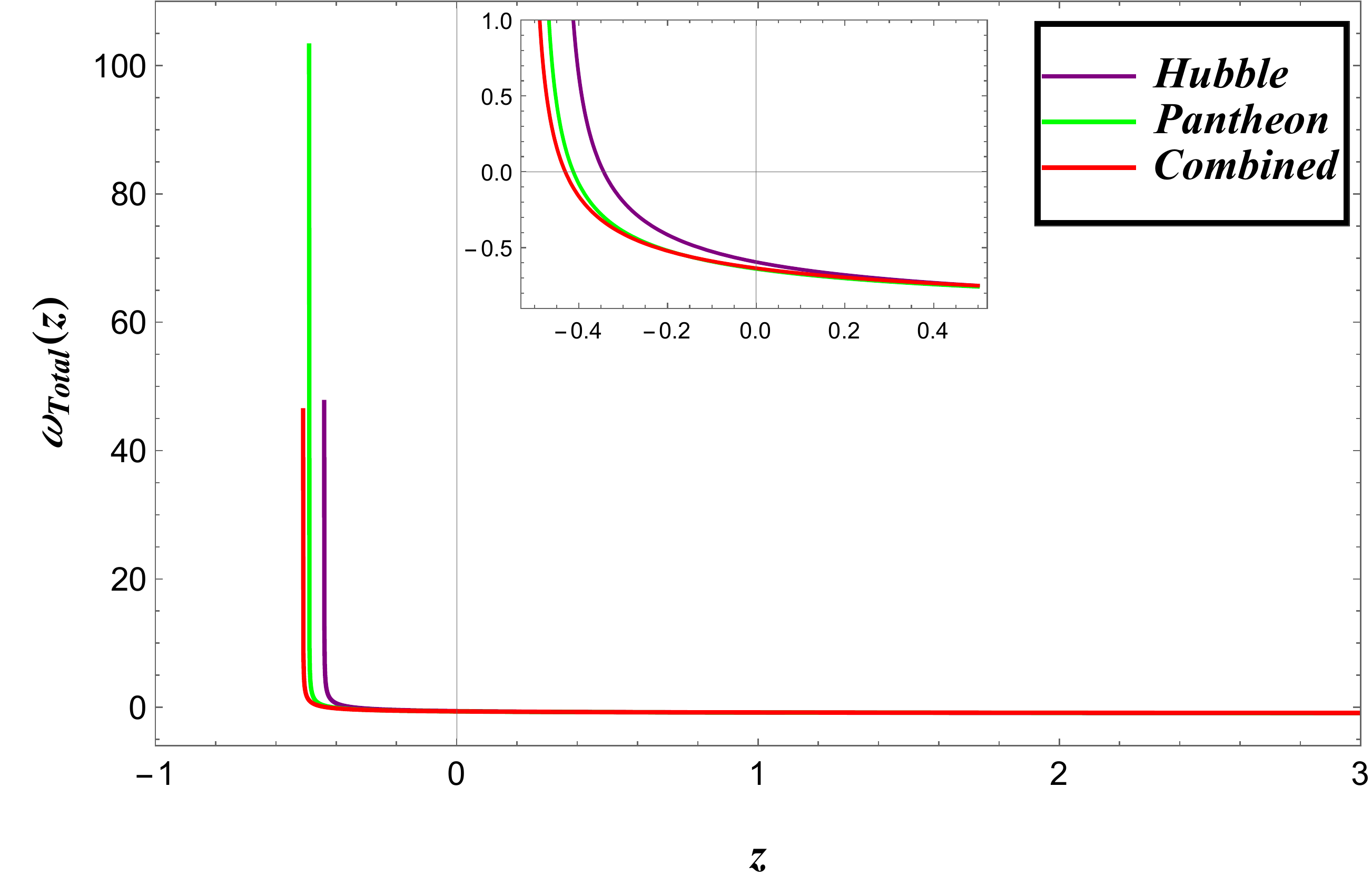} \\ 
\mbox (a) & \mbox (b)%
\end{array}
$%
\end{center}
\caption{{\protect\scriptsize The left panel (a) shows the evolution of EoS
parameter of dark energy $\protect\omega _{\protect\phi }$ \textit{vs.}
redshift }${\protect\scriptsize z}${\protect\scriptsize \ and the right
panel (b) shows the evolution of total EoS parameter $\protect\omega %
_{Total} $ \textit{\ vs.} redshift }${\protect\scriptsize z}$%
{\protect\scriptsize .}}
\label{EoS}
\end{figure}

In the Fig. \ref{EoS} the left panel (a) shows the redshift evolution of $%
\omega _{{\phi }}$, where it can be observed that the present value of $%
\omega _{{\phi }}$, $\omega _{{\phi 0}}$ for different observational
statistical datasets is $\approx -0.85$, which represents that $\omega
_{\phi }$ remains in quintessence region at present and is consistent with
Riess \cite{Reiss1998}. Fig (b) depicts the class of $\omega _{total}$ and
predicts its present evolution in the quintessence regime and becomes
positive $(\omega _{total}>0)$ in the near future scenario. This observation
reveals that Universe suddenly collapsed in the late times and led to Big
Crunch.

\subsection{Jerk parameter}

The evidence of a rapidly expanding Universe may be interpreted in a variety
of ways, including examining the behaviour of some higher derivatives of the
scale factor. The geometrical behaviour of the higher-order derivatives of
these cosmographic parameters ($a$, $H$, or $q$) allows us to investigate
the performance of a variety of dark energy models. Here, we try to intend
the behavior of third-order derivative of scale parameter i.e. jerk
parameter $j$ given as 
\begin{equation}
j=\frac{\dddot{a}}{aH^{3}}.  \label{jerk}
\end{equation}

In terms of redshift $z$, the jerk parameter $j$ can be written as, 
\begin{equation}
j(z)=-q+2q(1+q)+(1+z)\frac{dq}{dz}.  \label{jerk2}
\end{equation}

The following plot shows the evolution of the jerk parameter 
\begin{figure}[tbp]
\begin{center}
$%
\begin{array}{c@{\hspace{0.1in}}c}
\includegraphics[width=3.0 in, height=2.5 in]{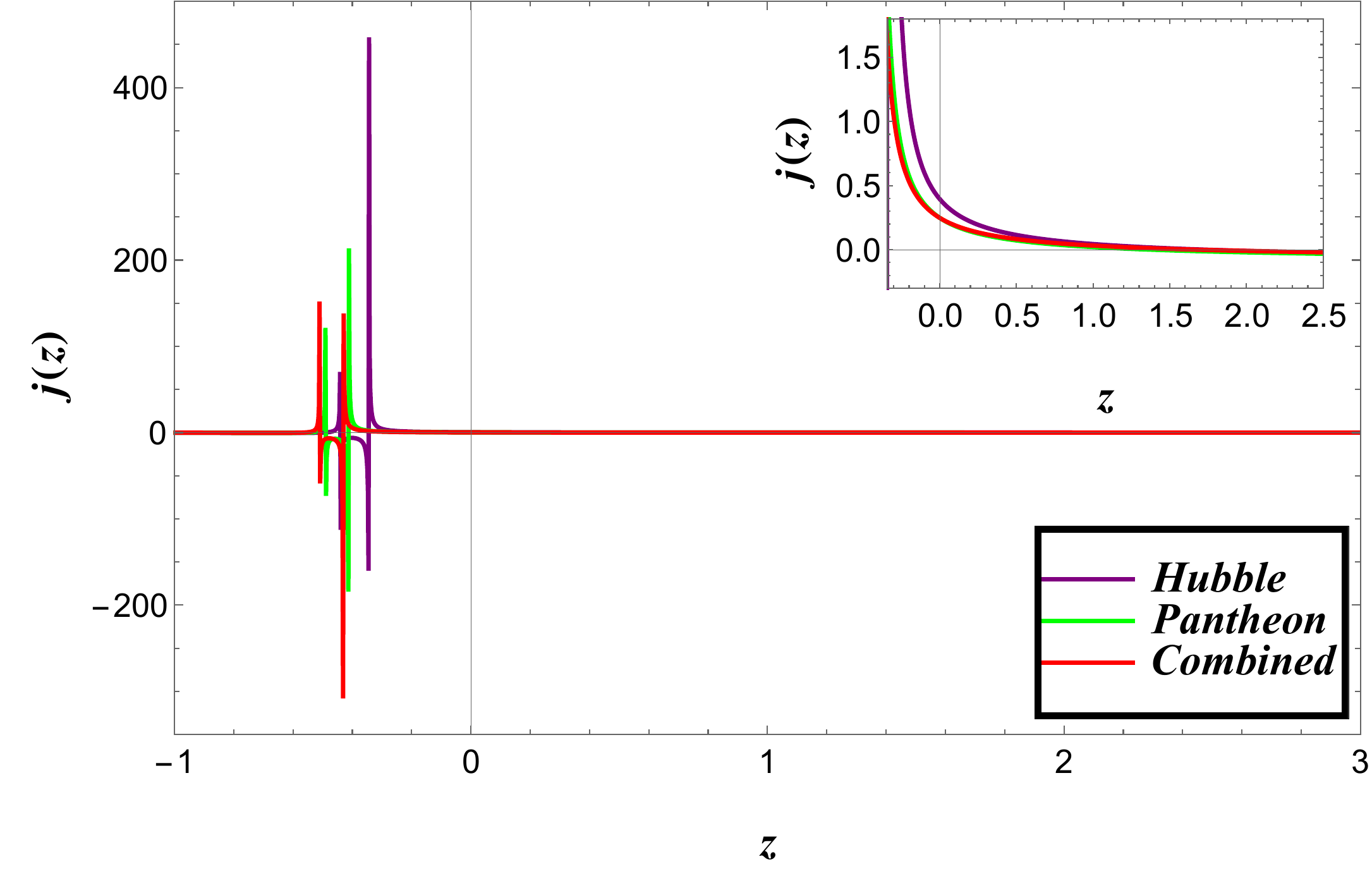} & 
\end{array}
$%
\end{center}
\caption{{\protect\scriptsize The evolution of jerk Jerk parameter $j$ 
\textit{\ vs.} redshift $z$.}}
\label{Jerk}
\end{figure}

From the above Fig. \ref{Jerk}, the plot of $j(z)$, it can be noticed that
initially, $j\approx 1$ and as time goes by, the value of $j$ decreases for
all three datasets. The current value of $j$ i.e. $j_{0}\approx 0$ for
Hubble dataset and for pantheon it is observed to be $j_{0}\approx 0.2$. The
obtained value of both datasets deviates from the value of the jerk
parameter, which is $j_{0}=1$ as per the standard $\Lambda CDM$ model. This
deviation of the observed and obtained value of $j$ suggests that our dark
energy model has contrasting aspects as to the standard $\Lambda CDM$ model.%
\textbf{\ }

\subsection{Statefinder diagnostic}

The Hubble parameter $H$ and the deceleration parameter $q$ are two of the
most ancient geometric parameters that describe the development of the
Universe. In order to distinguish the different dark energy models and
understand their behavior, few other geometric parameters are considered by
the theoreticians. The pair of geometric quantities known as statefinder
dignostic parameters $(s,r)$ and $(q,r)$ have been proposed by V. Sahni et
al. \cite{SR}, which are higher order derivatives of the cosmic scale factor
and are similar to that of $H$ and $q$. These parameters are defined as, $r=%
\frac{\dddot{a}}{aH^{3}}$, $s=\frac{r-1}{3(q-\frac{1}{2})}$ which can also
be represented in terms of redshift $z$ as,

\begin{equation*}
r=1-2\frac{1+z}{H}\frac{dH}{dz}+\frac{\left( 1+z\right) ^{2}}{H^{2}}\left( 
\frac{dH}{dz}\right) ^{2}+\frac{\left( 1+z\right) ^{2}}{H}\frac{d^{2}H}{
dz^{2}}
\end{equation*}

\begin{equation*}
s=\frac{-2\frac{1+z}{H}\frac{dH}{dz}+\frac{\left( 1+z\right) ^{2}}{H^{2}}%
\left( \frac{dH}{dz}\right) ^{2}+\frac{\left( 1+z\right) ^{2}}{H}\frac{d^{2}H%
}{dz^{2}}}{3\left( \frac{1+z}{H}\frac{dH}{dz}-\frac{3}{2}\right) }
\end{equation*}

The numerous trajectories in the $s-r$ and $q-r$ planes show the
chronological evolution of several dark energy concepts. Few fixed points in
these planes are $(s,r)=(0,1)$, $(q,r)=(-1,1)$ for $\Lambda $CDM model and $%
(s,r)=(1,1)$, $(q,r)=(0.5,1)$ for SCDM (standard cold dark matter) model.
The deviations of any dark energy model are then evaluated from these fixed
locations. For our obtained model, we have shown the $(s,r)$ and $(q,r)$
diagrams in the following plots.

\begin{figure}[tbp]
\begin{center}
$%
\begin{array}{c@{\hspace{0.1in}}c}
\includegraphics[width=3.0 in, height=2.5 in]{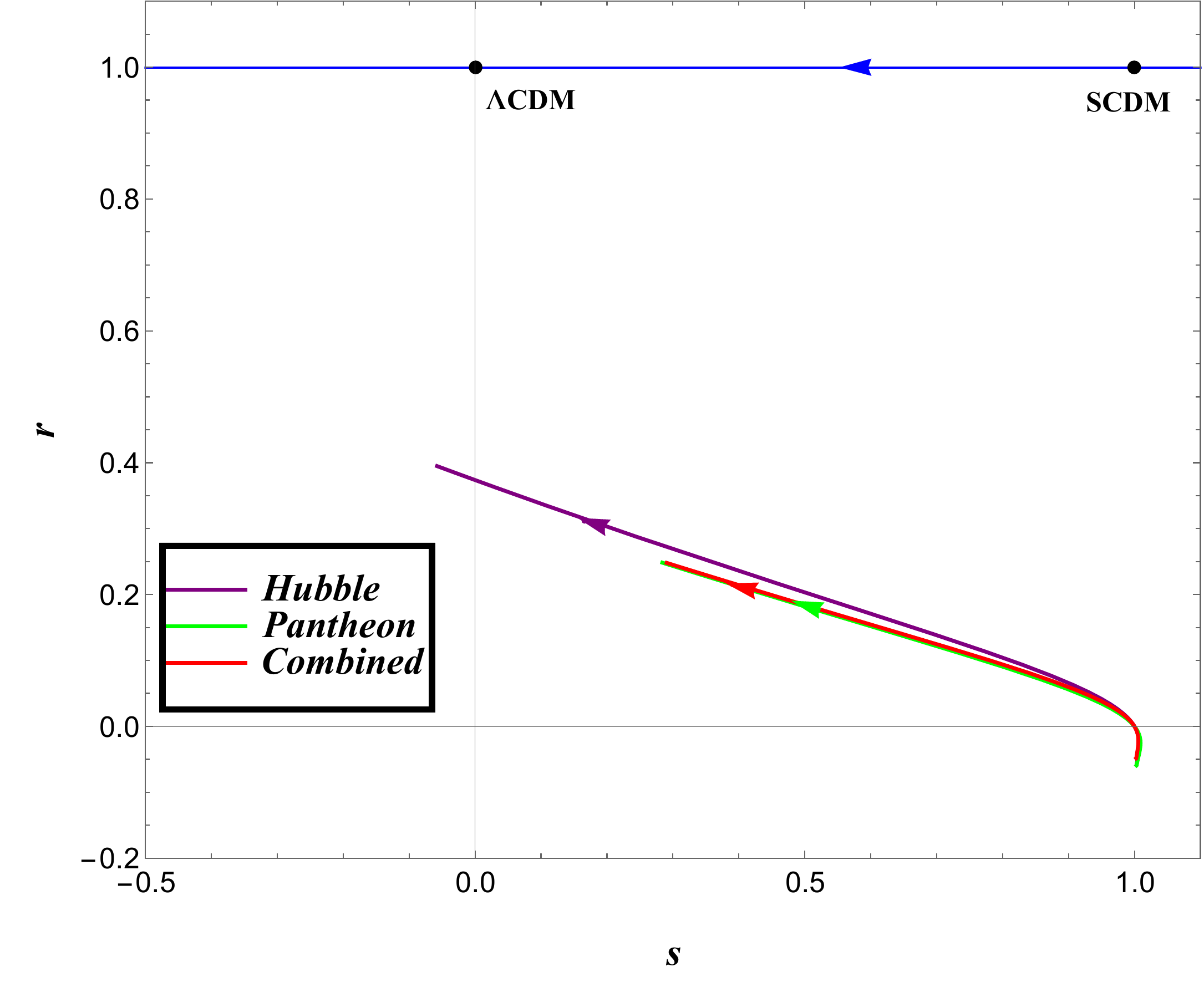} & %
\includegraphics[width=3.0 in, height=2.5 in]{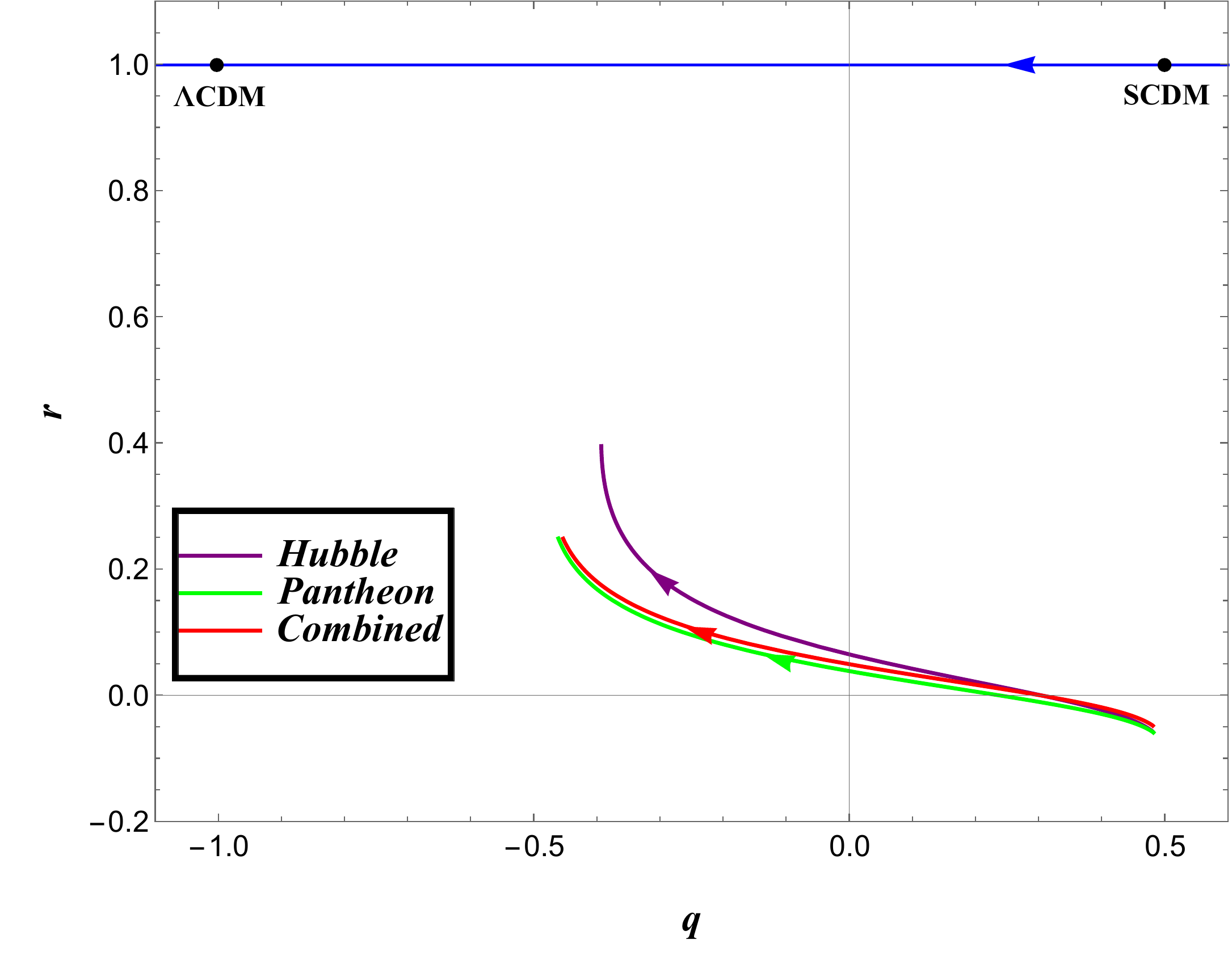} \\ 
\mbox (a) & \mbox (b)%
\end{array}
$%
\end{center}
\caption{{\protect\scriptsize In the figure the left panel (a) shows the }$%
{\protect\scriptsize s-r}${\protect\scriptsize \ plane and the right panel
(b) shows the }${\protect\scriptsize q-r}${\protect\scriptsize \ plane,
where the dark energy model shows its evolution.}}
\label{s-r}
\end{figure}

%

The above $(s,r)$ and $(q,r)$ diagrams shown in Fig. \ref{s-r} describe the
nature of the dark energy. As expected, the model has large deviation from
the $\Lambda $CDM model in the future.

\subsection{Om diagnostic}

A diagnostic tool known as Om diagnostic \cite{Om} has been utilized in
addition to several ways to quantify the contrast of the $\Lambda CDM$ model
among various DE models. This method discriminates between different DE
models without taking into account the EoS parameter $\omega $. It looks at
how different trajectories of $Om(z)$ behave with regard to redshift $z$ and
it can be assessed by considering $H$ \ \textit{w.r.t.} $z$. The form of $%
Om(z)$ is\textbf{\ } 
\begin{equation}
Om(z)=\frac{\{\frac{H(z)}{H_{0}}\}^{2}-1}{(1+z)^{3}-1}.  \label{40}
\end{equation}

For our model, the Om diagnostic analysis is shown in the following Fig. \ref%
{Omd}.

\begin{figure}[tbp]
\begin{center}
$%
\begin{array}{c@{\hspace{0.1in}}c}
\includegraphics[width=3.0 in, height=2.5 in]{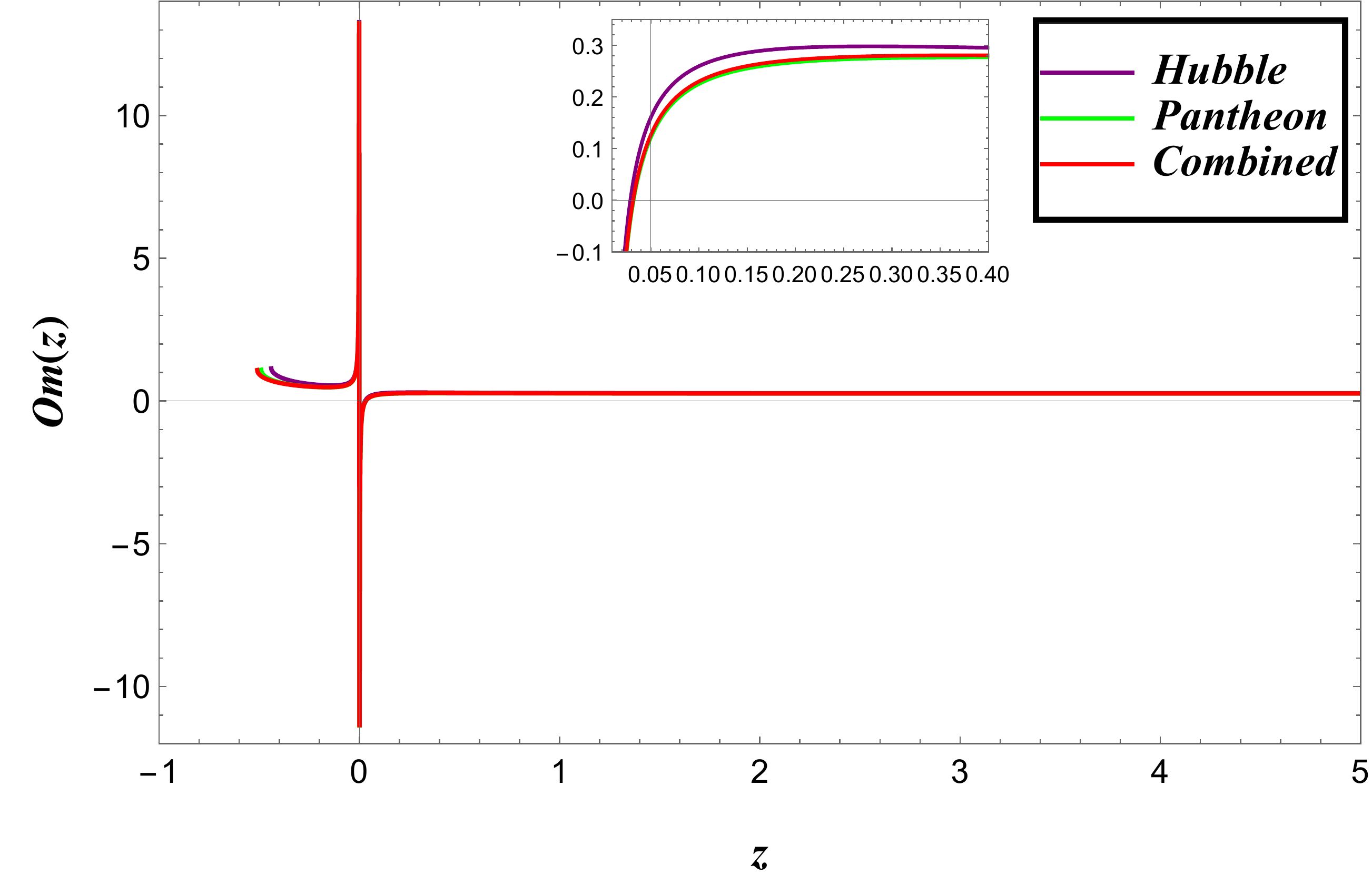} & 
\end{array}
$%
\end{center}
\caption{{\protect\scriptsize The plot of Om diagnostic $Om(z)$ \textit{\
vs. } redshift $z$.}}
\label{fig:Omd}
\end{figure}

The action of the DE model is decided by the slope of the function $Om(z)$.
The positive and negative curvature of $Om(z)$ w.r.t. $z$ represent phantom
and quintessence DE model. The zero curvature of $Om(z)$ represents the
standard $\Lambda CDM$ model. In the plot of $Om$ diagnostic, the
trajectories are initially increasing as redshift $z$ decreases, which means
that $Om$ has negative curvature, which corresponds to the quintessence dark
energy model. Once the trajectories attained the highest value, near to $%
z\approx 0$, the pattern of the trajectories suddenly falls to a negative
amount.

\subsection{Energy Conditions}

In GR, several conditions inhibit some regions where energy density becomes
negative to acquire the realistic model of the Universe. These conditions
are noticed as energy conditions. Many extensive applications of energy
conditions also serve to examine the viability of some crucial singularity
issues associated with the area of GR, warm holes, and black holes etc. The
standard energy conditions in $GR$ in terms of energy density $\rho $ and
pressure $p$ are specified as:\newline

\begin{itemize}
\item Weak energy condition (WEC) \textit{i.e.,} $\rho \geq 0 $, $\rho+p_i
\geq 0 $, where $i=1,2,3 $,

\item Null energy condition (NEC) \textit{i.e.,} $\rho+p_i \geq 0 $, where $%
i=1,2,3 $,

\item Strong energy condition (SEC) \textit{i.e.,} $\rho+\sum_{i=1}^3 p_i
\geq 0 $, $\rho+p_i \geq 0 $, where $i=1,2,3 $,

\item Dominant energy condition (DEC) \textit{i.e.,} $\rho \geq 0 $ , $\rho
\geq |p_i| $, $\forall i $, where $i=1,2,3 $ .
\end{itemize}

\begin{figure}[tbp]
\begin{center}
$%
\begin{array}{c@{\hspace{.1in}}cc}
\includegraphics[width=2.2 in, height=2.2 in]{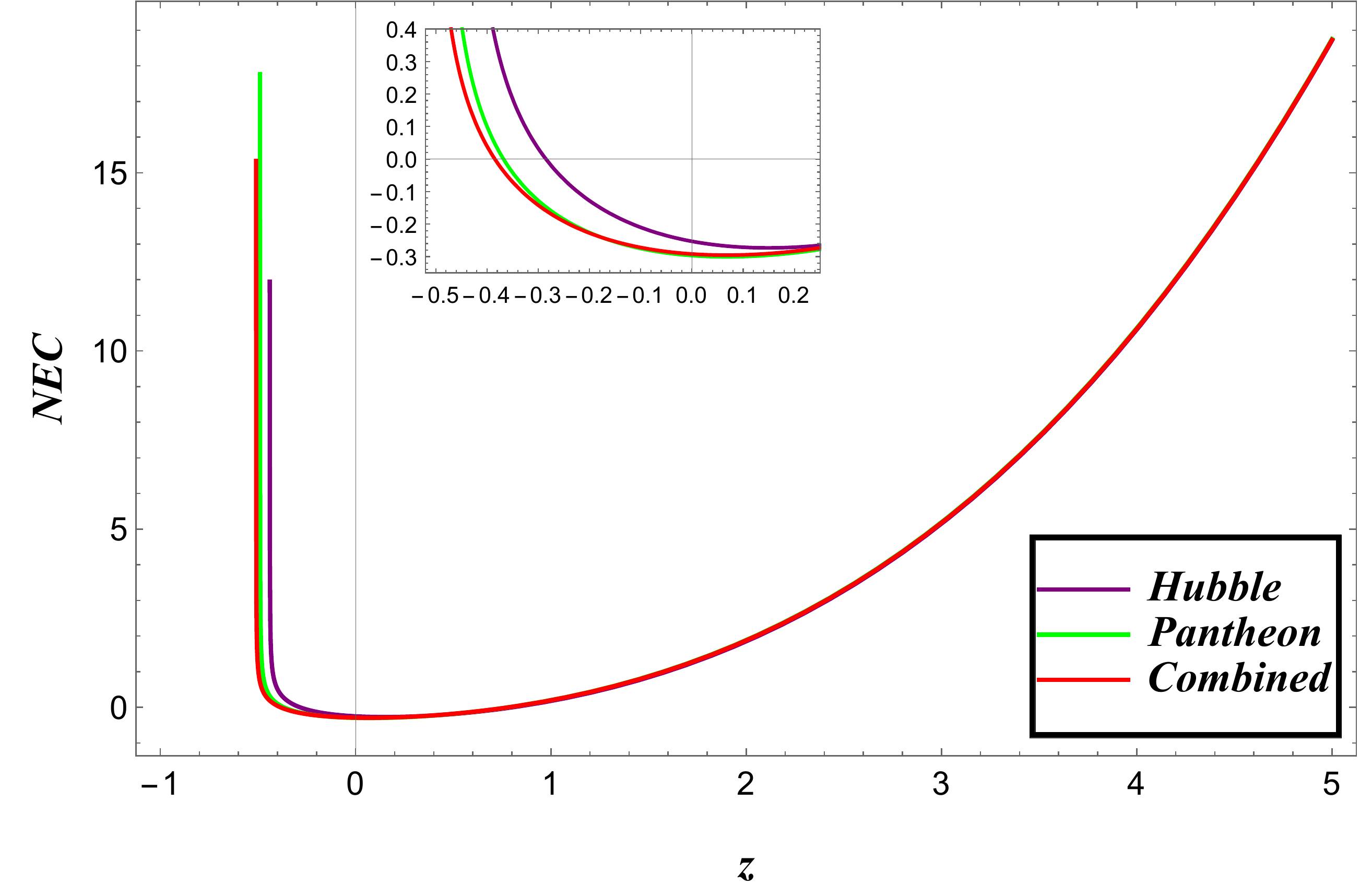} & %
\includegraphics[width=2.2 in, height=2.2 in]{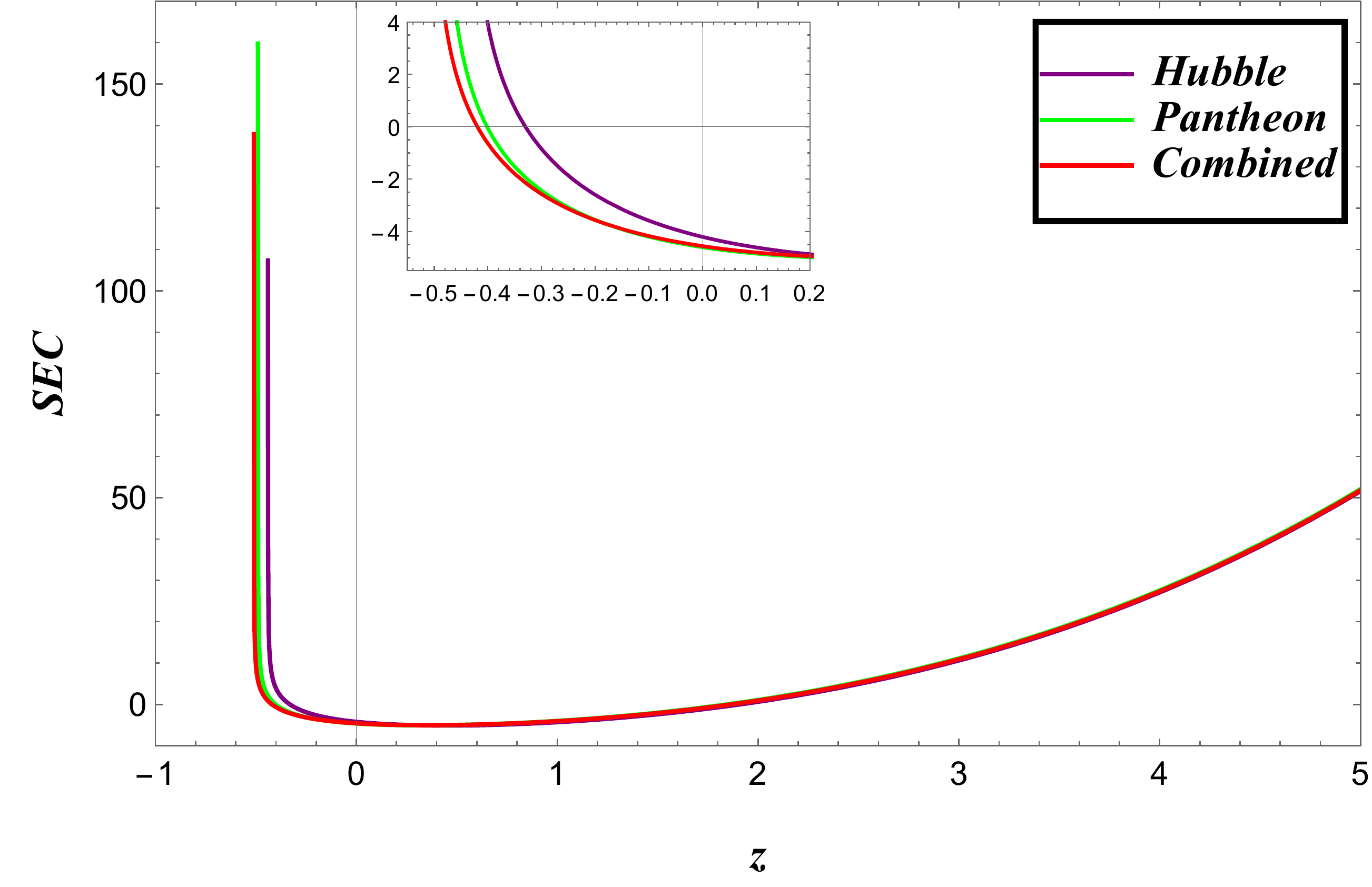} & %
\includegraphics[width=2.2 in, height=2.2 in]{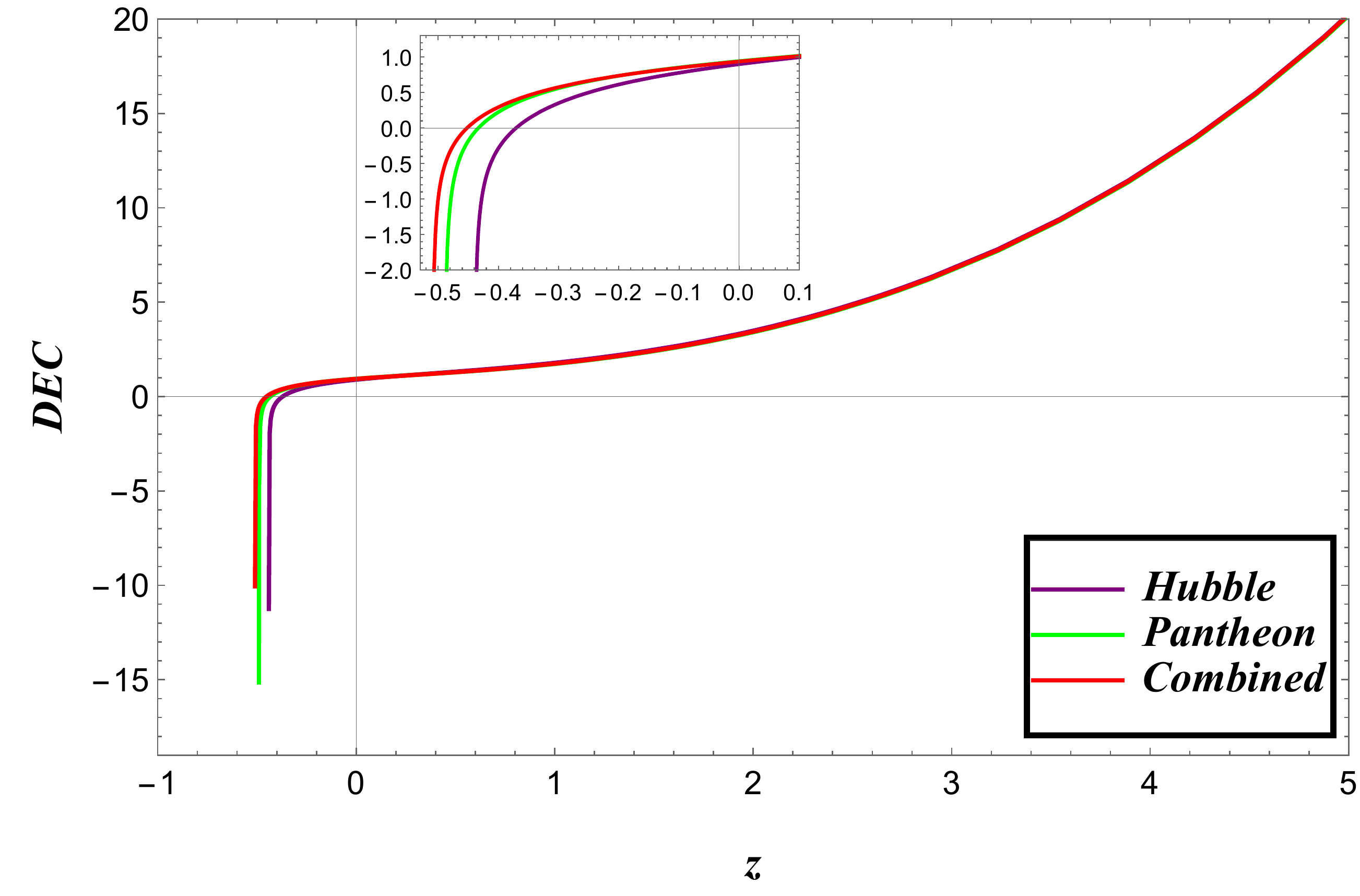} \\ 
\mbox (a) & \mbox (b) & \mbox (c)%
\end{array}
$%
\end{center}
\caption{{\protect\scriptsize The plots of energy conditions NEC, SEC and
DEC for the model.}}
\label{fig:EC}
\end{figure}

From the plots of energy conditions shown in Fig. \ref{EC}, one can notice
that null and dominant energy conditions are fascinated. Still, the strong
energy condition is turned off, corresponding to the Universe's accelerating
expansion. The achievement of null and dominant energy conditions on this
version is fairly the requisite requirement to impose conditions that make
the positive energy density in the Universe. However, the infraction of $%
\rho +3p>0$ suggests the presence of exotic matter in the Universe.

\section{Conclusion}

\label{sec7}

We examined a dark energy cosmological model in the general theory of
relativity in depth in this essay. As a generic scalar field, dark energy is
considered as a contender. To find a consistent solution, we have taken an
appropriate parametrization of density parameter of dark energy $\rho _{\phi
}$. The obtained results are considered interesting in many respects and
represent two smooth transitions from decelerating phase to an accelerating
one at redshift $z\approx 0.79$ in the recent past and then again from
acceleration to deceleration in the future at redshift $z\approx -0.29$. The
Universe's second phase transition causes the Universe to constrict and
collapse into a \textit{Big Crunch} singularity (see Fig. \ref{H-z & q-z}).
To examine the evolution in the early phase of the Universe and to study the
different stages in the cosmic evolution of the Universe, we have remolded a
vital cosmic parameter (EoS parameter) in terms of deceleration parameter in
FLRW background. External datasets with updated lists, such as $Hubble$, $%
Pantheon$, and $BAO$, were utilised to constrain the model parameters and
achieve the best fit values of model parameters in the functional form of
the parametrization approach employed here. The major goal of this work is
to investigate the dynamics of the current speeding cosmos using a scalar
field as a candidate of dark energy in the context of the FLRW metric. As a
result, the major findings of our model have been recorded as follows.

\begin{itemize}
\item To understand the cosmic evolution, the expressions of some important
cosmological parameters have been drafted as a function of redshift `$z$'
and shown them graphically. The obtained model is also compared with the $%
\Lambda $CDM model.

\item The plots in Fig. \ref{hubble}, Fig. \ref{pantheon} and Fig. \ref%
{hubpanthbao} shown are the $2-d$ contour plots showing best fit values of $%
\alpha ${\scriptsize , }$\beta ${\scriptsize , }$\Omega _{m0}${\scriptsize , 
}$\Omega _{\phi 0}$ obtained from emcee codes for the observational $Hubble$%
, $Pantheon$ and combined $Hubble+Pantheon+BAO$ datasets with{\scriptsize \ }%
$1-\sigma ${\scriptsize \ }and{\scriptsize \ }$2-\sigma $ \ errors.

\item The plots in Fig. \ref{h-z} and Fig. \ref{mu-z} ($H(z)\sim z$ and $\mu
(z)\sim z$) show the fitting of our model and compared with the $\Lambda $%
CDM model together with the error bars for the $57$ points and $1048$ of the
considered Hubble datasets and Pantheon datasets.

\item In the Fig. \ref{densities}, the evolution of energy densities of
matter, dark energy, and the pressure of dark energy are shown according to
the values of the model parameters obtained. We can see the distinct
behavior of the dark energy density and pressure from the standard lore. The
density vanishes in the future with the increase in pressure positively
showing the slowing down of expansion of the Universe and finally collapse
to a Big Crunch.

\item The plots in the Fig. \ref{EoS}(a) indicates the redshift evolution of 
$\omega _{{\phi }}$ and $\omega _{{total}}$. It has been observed that the
present value of $\omega _{{\phi }}$ for considered statistical datasets in
the range of $-0.85$ approximately, which enact that $\omega _{{\phi }}$ is
in the quintessence region at present and is consistent with Riess \cite%
{Reiss1998}. The Fig. \ref{EoS}(b), which predicts the evolution of $\omega
_{{total}}$ is negative at present and becomes positive in the future. This
observation reveals that Universe collapses in the future and led to Big
Crunch.

\item Fig. \ref{Jerk} depict that, the deviation of the observed and
obtained value of jerk parameter $j(z)$ suggests that our dark energy model
has contrasting aspects as to the standard $\Lambda $CDM model.

\item Fig. \ref{s-r} depict the nature of the dark energy model. The
obtained model is in the quintessence region at present and show a large
deviation from the standard $\Lambda $CDM model in the future as the model
is a collapsing one.

\item From the plot shown in Fig. \ref{Omd} of Om diagnostic, it can be
noticed that the trajectories are initially increasing as redshift $z$
decreases, which means that Om diagnostic has negative curvature, and
corresponds to the quintessence model of DE. Once the trajectories attained
the highest value, near to present time, the pattern of the trajectories
suddenly falls and approaches a negative value.

\item From the plots of energy conditions in Fig. \ref{EC}, one can notice
that NEC and DEC are fulfilled but SEC is not. The disagreement of SEC
corresponds to the accelerating cosmic expansion due to the presence of
exotic matter in the Universe.
\end{itemize}

After considering the aforementioned points, we can conclude that our
research reveals a cosmological model that is very impressed with the
approach of parametrization reconstruction of some physical parameters and
saves a reasonable domain of knowledge for understanding various
cosmological scenarios directly from the beginning of the Universe's
evolution. Without a doubt, using some observational datasets in this work
offers model parameters a more precise range, allowing for a more thorough
investigation of geometrical and physical aspects. However, the current
essay is merely a first step in apprehending the nature of dark energy.

\section*{Acknowledgments}

\end{document}